\documentclass[12pt,manuscript]{aastex}

\usepackage{float}
\usepackage{subfigure}
\usepackage{rotating}
\usepackage{amsmath}
\usepackage{calc}
\usepackage{amssymb}
\usepackage{amsfonts}
\usepackage{bm}
\makeatletter
\let\KV@Gin@viewport@org\KV@Gin@viewport
\define@key{Gin}{viewport}{  \begingroup
  \edef\x{\endgroup
    \noexpand\setkeys{Gin}{viewport@org={#1}}  }\x
}
\makeatother

\slugcomment{Submitted to ApJ}

\shorttitle{Achieving Consistent Doppler Measurements from \SDO/HMI Vector
  Field Inversions}
\shortauthors{Schuck, P. W. \textit{et al.}}

\newcommand{\CODERED}[0]{COADRED}
\newcommand{\tensorfont}[1]{\boldsymbol{\mathsf{#1}}}
\newcommand{\NGain}{N_{\mathrm{G}}}
\newcommand{\eig}{\tensorfont{e}}
\newcommand{\ee}{\mathsf{e}}
\newcommand{\EIG}{\tensorfont{E}}
\newcommand{\SC}{\boldsymbol{b}}
\newcommand{\BIAS}{\boldsymbol{\beta}}
\newcommand{\bias}{\boldsymbol{B}}
\newcommand{\GC}{\boldsymbol{g}}
\newcommand{\GAIN}{\boldsymbol{\gamma}}
\newcommand{\gain}{\boldsymbol{G}}
\newcommand{\NMF}{N_{\mathrm{MF}}}
\newcommand{\NLS}{N_{\mathrm{LS}}}
\newcommand{\NP}{N_{\mathrm{W}}}
\newcommand{\ND}{N_{\mathrm{D}}}
\newcommand{\NR}{N_{\mathrm{R}}}

\newcommand{\VR}{V_\mathrm{R}}
\newcommand{\VN}{V_\mathrm{N}}
\newcommand{\VW}{V_\mathrm{W}}
\newcommand{\REAL}[2]{\mathbb{R}^{#1\times#2}}
\newcommand{\COMPLEX}[2]{\mathbb{C}^{#1\times#2}}
\newcommand{\VT}{\mathcal{V}}
\newcommand{\VSDO}{\boldsymbol{V}_\mathit{SDO}}
\newcommand*{\Uel}{U}

\newcommand{\VLOS}[1]{\boldsymbol{V}_{\mathrm{LOS}\mhyphen{#1}}}
\newcommand{\vLOS}[1]{\boldsymbol{v}_{\mathrm{LOS}\mhyphen{#1}}}

\newcommand{\vlos}[1]{v_{\mathrm{LOS}\mhyphen{#1}}}
\newcommand{\abs}[1]{\left|#1\right|}
\newcommand{\NT}{N_t}

\newcommand{\Jacobian}{\boldsymbol{\mathcal{J}}}
\newcommand{\vv}{\boldsymbol{v}}

\newcommand{\x}{\mbox{\boldmath{$x$}}}
\newcommand{\xhat}{\boldsymbol{\widehat{x}}}
\newcommand{\yhat}{\boldsymbol{\widehat{y}}}
\newcommand{\zhat}{\boldsymbol{\widehat{z}}}
\newcommand{\ephi}{\boldsymbol{\widehat{\Phi}}}
\newcommand{\etheta}{\boldsymbol{\widehat{\Theta}}}
\newcommand{\er}{\boldsymbol{\widehat{r}}}

\newcommand{\B}{\mbox{\boldmath{$B$}}}
\newcommand*{\SDO}{\textit{SDO}}
\usepackage[normalem]{ulem}
\newcommand{\mhyphen}{\hbox{\sout{ }}}
\DeclareMathOperator{\Leg}{P}
\DeclareMathOperator{\median}{median}
\DeclareMathOperator{\grad}{\boldsymbol{\nabla}}
\DeclareMathOperator{\Pbar}{\overline{\Leg}}

\newcommand{\acm}{\mathcal{C}}
\newcommand{\Index}{n}
\newcommand{\IS}{\boldsymbol{I}}
\newcommand{\KLeig}{\tensorfont{d}}
\newcommand{\deig}{\boldsymbol{\epsilon}}
\newcommand{\KLEIG}{\tensorfont{D}}
\newcommand{\DeltaD}{\mathsf{D}}
\newcommand{\Nt}{N_t}
\newcommand{\Identity}{\tensorfont{I}}
\newcommand{\LAMBDA}{\boldsymbol{\Lambda}}
\newcommand{\ALPHA}{\boldsymbol{\alpha}}
\newcommand{\ACM}{\boldsymbol{\acm}}
\newcommand{\PHI}{\boldsymbol{\Phi}}

\begin{document}

\title{Achieving Consistent Doppler Measurements from \SDO/HMI Vector Field Inversions}

\author{Peter W. Schuck\altaffilmark{1}}
\email{peter.schuck@nasa.gov}

\author{Spiro Antiochos\altaffilmark{2}}
\email{spiro.antiochos@nasa.gov}

\and

\author{K. D. Leka\altaffilmark{3}}
\email{leka@cora.nwra.com}

\and

\author{Graham Barnes\altaffilmark{3}}
\email{graham@cora.nwra.com}

\altaffiltext{1}{Space Weather Laboratory, Code 674,
Heliophysics Science Division,
NASA Goddard Space Flight Center
8800 Greenbelt Rd.
Greenbelt, MD 20771}
\altaffiltext{2}{Heliophysics Science Division
NASA Goddard Space Flight Center
8800 Greenbelt Rd.
Greenbelt, MD 20771}
\altaffiltext{3}{NorthWest Research Associates
3380 Mitchell Lane
Boulder, CO  80301}

\makeatletter{}\begin{abstract}
  NASA's \textit{Solar Dynamics Observatory} is delivering vector field
  observations of the full solar disk with unprecedented temporal and spatial
  resolution; however, the satellite is in a highly inclined geostationary
  orbit. The relative spacecraft-Sun velocity varies by $\pm3$~km/s over a day
  which introduces major orbital artifacts in the Helioseismic Magnetic Imager
  data. We demonstrate that the orbital artifacts contaminate all spatial and
  temporal scales in the data.  We describe a newly-developed three stage
  procedure for mitigating these artifacts in the Doppler data derived from
  the Milne-Eddington inversions in the HMI Pipeline. This procedure was
  applied to full disk images of AR11084 to produce consistent
  Dopplergrams. The data adjustments reduce the power in the orbital artifacts
  by 31dB. Furthermore, we analyze in detail the corrected images and show
  that our procedure greatly improve the temporal and spectral properties of
  the data without adding any new artifacts. We conclude that this new and
  easily implemented procedure makes a dramatic improvement in the consistency
  of the HMI data and in its usefulness for precision scientific studies.
\end{abstract}


\makeatletter{}\section{Introduction}
The Helioseismic Magnetic Imager (HMI) aboard NASA's \textit{Solar Dynamics
  Observatory} (\SDO) produces full-sun vector magnetic field observations at
$1\arcsec$ resolution with a cadence of approximately 12 minutes. These data
represent an unprecedented opportunity to study time evolution of solar vector
magnetic fields on the spatial scales and time scales of active region
evolution. For the first time relatively pristine data are available that are
uncontaminated by the Earth's atmospheric seeing, which causes distortions
that often cannot be completely corrected by speckle reconstruction imaging or
adaptive optics. In principle, the HMI data can lead to major advances in
science understanding, because the energy and helicity transported though the
photosphere and into the corona can be determined by measuring plasma
velocities (optical flow/image motions) from a sequence of vector magnetograms
\cite{Schuck2005a,Schuck2006a,Schuck2008d}. While speckle reconstruction
imaging and adaptive optics can dramatically improve the local resolution of
the images, these techniques often do not preserve the relative distances
between solar structures from frame to frame which introduces large artificial
biases in velocity estimates. In contrast, \SDO{} represents a stable platform
with a known pointing, located outside Earth's atmosphere, thereby potentially
permitting highly accurate measurements of the velocities between photospheric
features from frame to frame.\par
\SDO{}, however, is in a highly inclined geosynchronous orbit chosen so as to
maximize data throughput to the ground based receiving
stations. Unfortunately, this orbit produces a large $\pm$3~km/s variation in
the relative velocity between the HMI instrument and the Sun, which leads to
major orbital artifacts in the HMI data. Since the orbit is accurately known,
it would seem that removal of the artifacts should be straightforward; but,
even after five years into the mission, the exact mechanisms that contaminate
the data remain a mystery and the rigorous removal of the artifacts has not
been accomplished. There is speculation that the artifacts are caused by the
motion of the Fe~I $\lambda_0=6173.343\,$\AA{} line across the HMI
transmission filters as the satellite executes its geosynchronous orbit. Over
the period of an orbit the radial velocity of the satellite varies by as much
as $\pm3.2$~km/s, which corresponds to a shift of
$\Delta\lambda\simeq\lambda_0\,2\,\Delta{v}/c\simeq131\,$m\AA{}, which is
larger than the nominal HMI filter separation of
$\Delta\lambda\simeq69\,\mbox{m}$\AA. It is now well established that these
artifacts contaminate many observables computed from HMI data (\textit{e.g.},
Section~7.1 in \cite{Hoeksema2014}, Fig.~4 in \cite{Liu2012} for AR11072 shows
oscillations in the shear helicity flux, Fig.~3 in \cite{Chintzoglou2013}
exhibits clear dips in the magnetic field near midnight on 14 Feb, 15 Feb and
16 Feb, and the Poynting flux in Fig.~2 of \cite{Vemareddy2015} shows a very
clear 12~hr oscillation). The optimal solution would be to understand the
source of the contamination at the spectral level and correct optical
distortions though calibration prior to any inversion process to estimate
Doppler velocities and magnetic fields. However, even if the source of the
contamination is definitively identified and corrected for future
observations, it is unclear whether these corrections could be implemented for
the data already archived. If the HMI data are ever to be used for
high-precision unbiased studies or for data-driven modeling, it is absolutely
essential that a rigorous procedure be developed for mitigating the systematic
errors in the archived and future data.\par
In this paper we present a new approach \CODERED{}: Cleansing Orbital
Artifacts \--- Demodulation by REnormalizing Data for correcting the
down-stream Doppler velocities derived from the Milne-Eddington inversions. It
should be emphasized that all spatial and temporal scales are affected by the
artifacts, thereby rendering the data essentially useless for detailed
quantitative spatio-temporal analysis at the cadence of the data series. The
goal of our procedure describe below is to provide a data set that is
consistent, i.e., free of orbital artifacts. Note that we cannot claim to
derive data that are absolutely correct, because there are no absolute
calibrations for the measurements. Potentially inter-calibration between
vector magnetographs could determine which measurements are correct, but
presently there are no absolutely calibrated magnetographs to compare
with. Furthermore, even understanding the relative calibrations between
magnetographs is rife with complexities, because magnetograms often use
different magnetically sensitive lines corresponding to different heights in
the solar atmosphere, with different spatial resolutions, different spectral
sampling, and different time-cadence
\cite[]{Leka2001,Leka2009,Orozco2007,Leka2011,Leka2012,Leka2013}. Given these
many obstacles to determining absolute calibrations for the purposes of
``correcting'' any of the observables from the HMI Pipeline data we focus our
attention instead on producing \textit{artifact-free, consistent} data by
renormalizing the observables to the rest velocity of the satellite
$\VR\rightarrow0$. We emphasize that even the measurements
\underline{observed} at $\VR=0$ may not be ``correct'' in an absolute
sense. Indeed, we will show that significant biases remain. However, this
renormalization does remove most of the contamination correlated with the
spacecraft orbital velocity. The result is a rigorous consistent data that can
now be used for spatio-temporal analysis of the image dynamics.\par
\subsection{The \CODERED{} Procedure}
\CODERED{} consists of a three stage process for obtaining consistent
Doppler measurements\footnote{ Specifically the
  hmi.ME\_720s\_fd10\{vlos\_mag\} data.} with vector field data from the HMI
Pipeline. \\ {\bf 1.}  The first step is to remove the projection of the
satellite velocity along the line-of-sight (LOS) from each pixel. Since the
satellite orbit is known with high accuracy, then in principle, if the HMI
measurements were precise, there would be no orbital effects remaining in the
data. As will be shown below, however, this is definitely not the
case. \\ {\bf 2.} We next remove the three well-known quasi steady-state
signals from each image: the differential rotation, the meridional flows, and
the convective blue-shift. Note that the first two are actual physical flows,
but the latter is not. As discussed in detail by \cite{Beckers1978}, the
convective blue-shift, commonly referred to as the limb shift, is due to the
observed strong correlation in photospheric lines between intensity and
wavelength shift. For unresolved convective flows this correlation introduces
a systematic bias to any line shift determinations, and this bias has a strong
center-to-limb variation.  The three large-scale signals are removed by
fitting each image with a series of eigenfunctions representing the
differential rotation, the meridional flows, and the limb shift ``flows'', and
subtracting these from each image.  Not surprisingly, these biases vary with
\SDO{}'s orbital velocity, and thus by removing them from each image, we
reduce the power in the orbital effects substantially. We find that a low
order, $\sim 8$, series for each type of flow is sufficient for fitting and
removing the large-scale biases.\\ {\bf 3} After removal of these biases, we
expect that the only remaining physical effects in the residual images are the
small-scale convective dynamics, which should be largely
quasi-stationary. However we show below that these residual images still
exhibit substantial orbital artifacts. We conjecture that these artifacts are
caused by some type of interference between the instrument response and
convective structures moving across the solar disk, which produces spectral
artifacts by spatio-temporal modulation of the convective amplitudes. If so,
then we expect there to be a strong correlation of \SDO{}'s radial velocity
with the limb shift effect in the instrument response.  To remove this
artifact, we calculate the magnitude of each residual image and then fit this
with an eighth-order series of limb shift eigenfunctions. This yields the
dependence of the coefficients on the satellite radial velocity $\VR$. As will
be shown below, the coefficients exhibit a clear systematic dependence with
$\VR$, which must be due to orbital artifacts. Consequently, we simply
renormalize each pixel so that each image appears to be observed at the same
orbital velocity, \textit{arbitrarily} chosen in this case, to be
$\VR\equiv0$. This renormalization almost completely eliminates the orbital
signal, so then we simply add back in the large-scale flows to obtain HMI
Dopplergrams that exhibit essentially no orbital artifacts and that now can be
used for accurate science investigations.\par

In the Sections below we describe the data set that we used for this
analysis and describe exactly how the \CODERED{} procedure was applied 
to this data set. The data and the procedure are described in sufficient detail
that others can use our methods or can modify them for application to other
types of data sets.

\makeatletter{}\section{Data Description}
Approximately 17 days of \SDO{}/HMI data in 2010 are considered in developing
the \CODERED{} procedure. During this time period, the Sun was fairly quiet
but these days encompass the disk passage of the small and nearly potential
active region 11072 discussed by \cite{Liu2012}. This time-period was
specifically chosen for two reasons. First, it was relatively quiet period,
and therefore most pixels represented the same physical process of solar
convection. This permitted the straight-forward disentanglement of the orbital
effects and the physical solar effects. Second, there was coverage by both HMI
and MDI during this time.\par
The HMI has two independent cameras that produce several data series, which
undergo different analysis.  Consequently, these data sets may be used to
inter-compare observations of the Sun. Three data series are produced by the
vector field ``side camera'' (Keyword CAMERA=1). The
hmi.M\_720s\{Magnetogram\} and hmi.V\_720s\{Dopplergram\} series are produced
by an MDI-like algorithm \cite[]{Couvidat2012a} and the hmi.ME\_720s\_fd10
series with segments \{field, azimuth, inclination, vlos\_mag, etc\}
\cite[]{Hoeksema2014} which are produced by the Very Fast Inversion of the
Stokes Vector \cite[VFISV,][]{Borrero2009,Centeno2014} Milne-Eddington
code. Two data series are produce by the Doppler Camera (Keyword CAMERA=2)
hmi.M\_45s\{magnetogram\} and hmi.M\_45s\{Dopplergram\}
\cite[]{Schou2012,Scherrer2012,Hoeksema2014}.  These four series have
corrections and calibrations that are not applied to the spectral data
provided to the Mile-Eddington inversions which are part of the
hmi.ME\_720s\_fd10 series with segments \{field, azimuth, inclination,
vlos\_mag, etc, etc\} \cite[]{Hoeksema2014}. Furthermore, during 2010 there
was a significant period of overlap in the observing programs of HMI and
MDI. The MDI instrument produces two series which are comparable to
observables estimated by HMI namely mdi.fd\_M\_96m\_lev182 which is a LOS
magnetogram and mdi.fd\_V which is a LOS Dopplergram. Thus, in principle,
there are three independent cameras and 4 different data sets for each LOS
observable available for inter-comparison for AR11084. This paper will focus
on the hmi.ME\_720s\_fd10\{vlos\_mag\} data which with vector field ``side
camera'' which measures 4096x4096 filtergrams at six wavelengths and four
polarizations of the \ion{Fe}{1} 617.3~nm line which are corrected for solar
rotation, cosmic rays, distortions and other effects \cite[]{Hoeksema2014}.
These spectral data are used to construct the Stokes parameters (I,Q,U,V)
which are in turn inverted based on the Milne-Eddington approximations using
VFISV (Very Fast Inversion of the Stokes Vector) to produce about 10 physical
parameters including the magnetic components relative to the line of sight
$\B_\xi$, $\B_\eta$, $\B_\zeta$ and the magnetized plasma velocity along the
line of sight $v_{\mathrm{LOS}}$. Since the data are spectrally sparse and the
Milne-Eddington model is a simplified description of the line-forming physics,
the inversion of these spectra is highly sensitive to the relative velocity
between the instrument and the line forming region on the Sun.  \par
\subsection{Camera\#1 Orientation}
The pointing of HMI is known to very high accuracy due to the Venus transit
data of June 5-6, 2012 \cite[]{Couvidat2014,Emilio2015}.  In particular, The
CROTA2 keyword is often taken to be the orientation of the Sun's North pole in
Solar images. However, this interpretation depends specifically on how CROTA2
is determined. For HMI, the CROTA2 keywords is known to better than
$0.002^\circ$ \cite[]{Couvidat2014} \textit{relative to the transit of
  Venus}. We caution the reader that this is not an absolute determination
relative to the Sun's North Pole either relative to magnetic phenomena or
solar flows.  Furthermore, the P-angle and B0-angle estimates in the HMI
Pipeline keywords do not include the corrections to the Carrington elements
determined by \cite{Beck2005} using time-distance helioseismology on data from
\textit{SOHO}/MDI. \cite{Beck2005} found that Carrington's estimates of $i$
the angle between the plane of the ecliptic and the solar equator, and
$\Omega$ the angle between the cross point of the solar equator with the
ecliptic and the vernal equinox were off by as much as $\Delta
i=0.095\pm0.002^\circ$ and $\Delta\Omega=-0.17\pm0.1^\circ$ with the error in
$i$ effecting the error in P-angle and CROTA2 more than the error in $\Omega$.
This may introduce a temporal error in CROTA2 of as much as $0.1^\circ$
depending on the heliocentric ecliptic longitude of the \SDO. Furthermore,
\cite{Beck2005} concluded that the cross equatorial flow they measured could
correspond to a systematic bias in P-angle of another $\simeq0.1^\circ$ under
the assumption that the Sun does not maintain long term cross equatorial
flows. These results suggest that generally there may be a time-dependent
error in the inferred direction of Solar North of as much as
$\simeq0.2\--0.3^\circ$ when using the standard Carrington elements (This is
not an HMI specific statement). Knowing these systematics is important for
interpreting meridional flows as we shall see in Section~\ref{sec:stage2}.\par
\begin{figure}[!t]
\begin{center}
\includegraphics[width=6in,viewport=20 293 700 517,clip=]{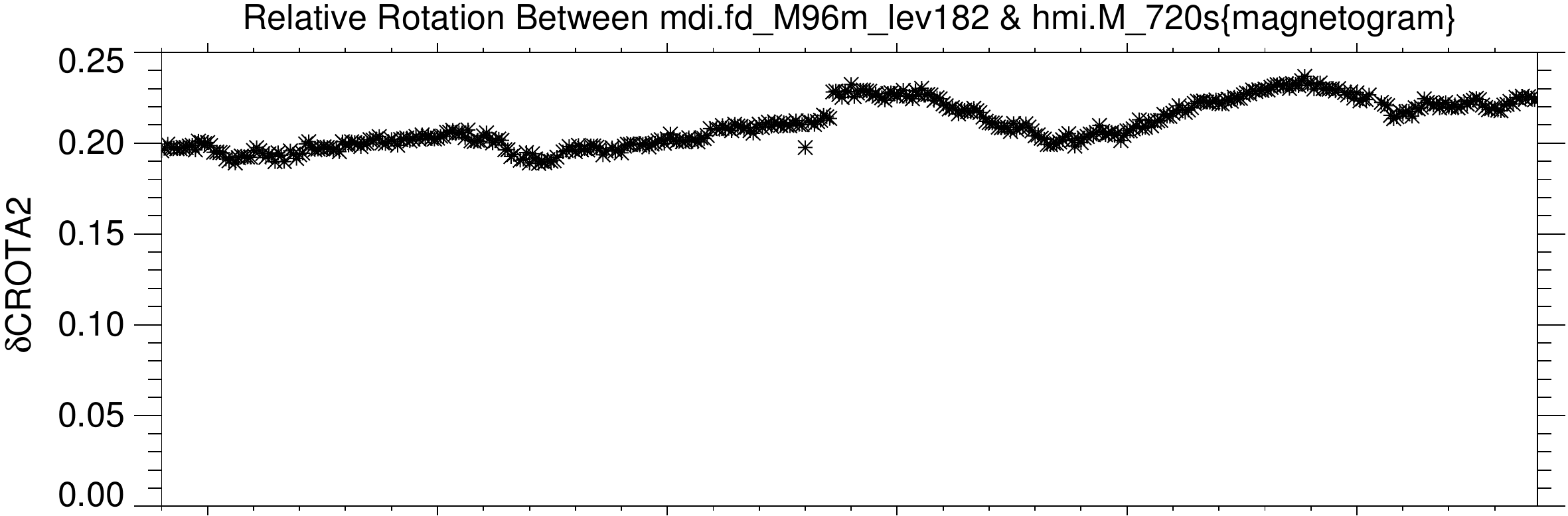}
\includegraphics[width=6in,viewport=20 38 700 73,clip=]{f1}
\caption{Relative roll-angle between HMI Camera\#1 and MDI based on the
  hmi.M\_720s series.\label{fig:mdi_roll}}
\end{center}
\end{figure}
Figure~\ref{fig:mdi_roll} shows the relative angle between HMI Camera\#1 and
MDI based on the normalized cross-correlation of the mdi.fd\_M\_96m\_lev182
and the hmi.M\_720s series during the 2010 June. The MDI series keywords
CRVAL1 and CRVAL2 were adjusted by 1.9 and -0.5 respectively to maximize the
cross-correlation coefficient $C\simeq0.9$ for the results presented here. The
MDI image was convolved with a Gaussian filter, $\sigma_\mathrm{MDI}=1$~pixel,
and the HMI image was convolved with
$\sigma_\mathrm{HMI}=4\times0.875=3.5$~pixels, chosen to maximize the
cross-correlation. Convolving both data sets with a Gaussian reduces shot
noise. The CROTA2 value for the HMI image was then adjusted by
$\mbox{CROTA2}\rightarrow\mbox{CROTA2}-\delta\mbox{CROTA2}$ over a range of
$-1^\circ$ to $1^\circ$ where positive $\delta\mbox{CROTA2}$ corresponds to a
counterclockwise(clockwise) rotation of HMI(MDI) to bring them into
alignment. The Carrington/Stonyhurst coordinates were used to determined the
same locations on each image and the blurred HMI image was re-sampled to MDI
observations. The offset between the two cameras lies between $0.18^\circ$ and
$0.23^\circ$. The variation could be due to a drift in CRPIX or roll in
MDI. At present, there is no predictive or definitive attitude data for SOHO
during this time
period,\footnote{\url{http://sohowww.nascom.nasa.gov/data/ancillary/\#attitude}}
but these values are in rough agreement with \cite{Liu2012a} who found
$\delta\mbox{CROTA2}\simeq0.22-0.10=0.12^\circ\pm0.05^\circ$ in a comparison
between the same two data series using a slightly different procedure. These
results imply that a $\mbox{CROTA2}$ value of 180 corresponds closely to
solar North pointing downward in the images to within a few tenths of a
degree.\par
\subsection{Co-Registration}
\begin{figure}[!t]
\centerline{\includegraphics[width=4in]{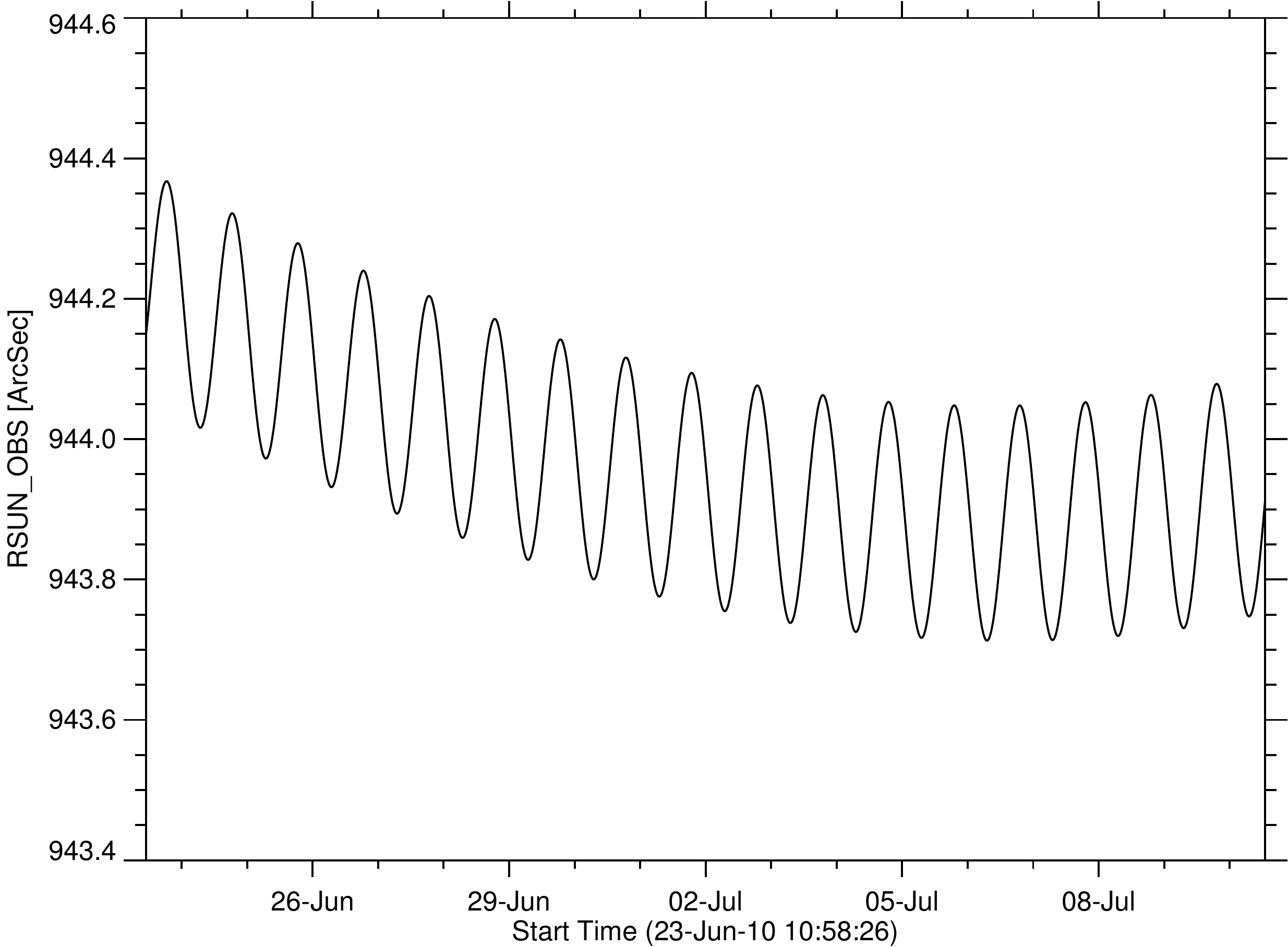}}
\caption{The radius of the Sun in arc-seconds as observed by \SDO/HMI. Since
  the resolution of each pixel is fixed in the images, this implies that the
  effective solar radius in pixel units is varying from image-to-image in the
  hmi.ME\_720s\_fd10 series as \SDO{} orbits the earth.  \label{fig:f2}}
\end{figure}

Figure~\ref{fig:f2} shows the radius of the Sun as observed by HMI over the
17 day period. On the time scale of a day, the diameter(radius) of the Sun
varies by roughly $1\arcsec\left(0.5\arcsec\right)$ corresponding to roughly
two pixels and on the time-scale of 17 days the diameter(radius) of the Sun
varies by roughly $2\arcsec\left(1\arcsec\right)$ corresponding to roughly
four pixels. For spatio-temporal analysis, the optimal solution would be to
track the active region across the Sun at the average speed of the active
region. However, non-uniformities in the HMI instrument response then convolve
noise in space and time (see movies). 
The Co-registration choice for this investigation is to: (1) shift each image
so that the center of the Sun corresponds to the center of the image
$\left(\mbox{CRPIX1}=2048.5,\mbox{CRPIX2}=2048.5\right)$, (2) rotate each
image to the same orientation $\left(\mbox{CROTA2}=0\right)$, and remap each
image to the same observation point $\mbox{DSUN\_OBS}=152017949201$~m
(distance to the Sun), so that the solar radius remains constant in pixel
units. Using this co-registration convention each pixel corresponds to the
same nominal location on the solar disk.\par
 
\makeatletter{}\section{Doppler Data Processing}
The following sections describe the exact procedure by which the HMI data
are processed in order to remove the orbital artifacts. In the subsequent
analysis the variable $U$ is used to represent components of a
velocity on the surface of the Sun and and the variable $\VSDO$ is
used the spacecraft velocity, but generally $\VT$ is used to represent
a theoretical model of the line-of-sight (LOS) Doppler velocity and $v_\mathrm{LOS}$
is used to represent various stages of data processing $\vlos{0}$,
$\vlos{1}$, $\vlos{2}$, and $\vlos{3}$ with $\vlos{0}$ representing
the raw observed Doppler velocity form the HMI Pipeline. The
coordinate systems used are described in the Appendices based on the
notation in \cite{Thompson2006}. In particular,
Appendix~\ref{sec:Stony} describes the projection of the Stonyhurst
unit vector $\left(\ephi,\etheta,\er\right)$ onto the LOS.
\subsection{Stage~1: Subtraction of the LOS Projection of Satellite
  Velocity\label{sec:satellite:v}} 
The first stage in the analysis is the
removal of the LOS projection of the satellite velocity
(Sat-V) from each pixel.  The Heliocentric-Cartesian LOS direction
$\left(\xhat,\yhat,\zhat\right)$ in helioprojective coordinates
$\left(\theta_\rho,\psi\right)$ is (See Appendix~\ref{app:observer})
\begin{equation}
\boldsymbol{\widehat{\eta}_\mathrm{LOS}}\left(\theta_\rho,\psi\right)=\sin\theta_\rho\,\sin\psi\,\xhat-\sin\theta_\rho\,\cos\psi\,\yhat+\cos\theta_\rho\,\zhat.\tag{\ref{eqn:eta:los}}
\end{equation}
Positive values of Doppler shift correspond to (redshifts) motion away from
the
satellite.\footnote{\url{http://jsoc.stanford.edu/doc/data/hmi/sharp/old/sharp.MB.htm}}
In principle, the LOS plasma motion on the Sun can be determined by
subtracting the projection of the satellite motion onto the LOS from the
measured Doppler velocity at each pixel
\begin{equation}
\vlos{1}=\vlos{0}-\boldsymbol{\widehat{\eta}_\mathrm{LOS}}\cdot\VSDO\equiv-\boldsymbol{\widehat{\eta}_\mathrm{LOS}}\cdot\boldsymbol{\Uel}_\mathrm{surface}.\tag{\ref{eqn:satellite}}
\end{equation}
\begin{figure}[!h]
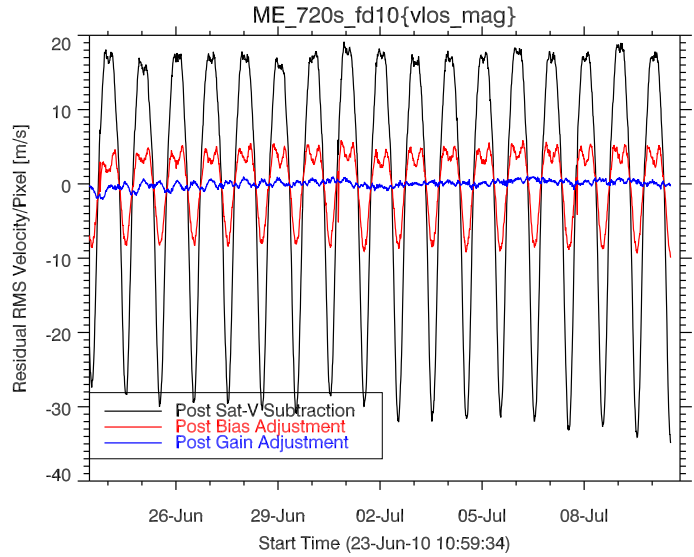

\centerline{\includegraphics[width=4in]{{{f3}}}}
\caption{Residual spatially averaged RMS Doppler velocity during 17 days in 2010. (Black) after the projection of the satellite velocity has been
  removed from each pixel. (Red) after solar rotation $V_\mathrm{Rot}$,
  meridional flows $V_\mathrm{MF}$, and limb shift $V_\mathrm{LS}$, have been
  removed.  \label{fig:raw_doppler}}
\end{figure}
The velocity vector for \SDO{} in the heliocentric Cartesian directions is
\begin{subequations}
\begin{align}
\VSDO=&\left(\VW,\VN,\VR\right),\\
=&\left(\mathrm{OBS\_VW},\mathrm{OBS\_VN},\mathrm{OBS\_VR}\right),
\end{align}
\end{subequations}
in the Heliographic Cartesian coordinate system and thus
\begin{equation}
\boldsymbol{\widehat{\eta}_\mathrm{LOS}}\cdot\VSDO=\mathrm{OBS\_VW}\,\sin\theta_\rho\,\sin\psi-\mathrm{OBS\_VN}\,\sin\theta_\rho\,\cos\psi+\mathrm{OBS\_VR}\,\cos\theta_\rho.\label{eqn:eta}
\end{equation}
Here $\VR$ corresponds to the HMI keyword ``OBS\_VR'' which is the
velocity of the observer in radial direction (positive is away from
Sun). Additionally $\VW$ corresponds to the the HMI keyword ``OBS\_VW'' which
is the velocity of the observer solar-westward (positive in the rough
direction of Earth orbit), and $\VN$ corresponds to ``OBS\_VN'' which is the
velocity of the observer solar-westward (positive in the direction of solar
north).\par
If the Doppler velocities are measured accurately, the $\vlos{1}$
would show no correlation with satellite motion. The black line in
Figure~\ref{fig:raw_doppler} shows the residual spatially averaged RMS Doppler
velocity
\begin{equation}
\mathrm{Residual}=\left\langle\Delta\vv_\mathrm{LOS}^2\right\rangle_x-\left\langle\Delta\vv_\mathrm{LOS}^2\right\rangle_{x,t},
\end{equation}
during 17 days in 2010 after the projection of the satellite
velocity has been subtracted from each pixel via~(\ref{eqn:satellite}). The
subscripts ``$x$'' and ``$t$'' indicate spatial and temporal averaging
respectively.  A large, daily oscillation remains indicating
\textit{systematic} errors in the LOS Doppler measurements from the vector
field inversions. Similar oscillations are known to occur in the $B$ from the
vector field inversions \cite[see][]{Hoeksema2014} and in the
hmi.V\_720s\{Dopplergram\} series from the same camera and in the 45~s cadence
data from the hmi.V\_45s series observed by camera 2 (private communication
with Phil Scherrer). This strong fixed oscillation convolves with and pollutes all spatial and temporal
scales, which greatly diminishes the usefulness of the data for science studies. 
\subsection{Stage~2: Large Scale Biases\label{sec:stage2}}
The second stage in the \CODERED{} procedure is the decomposition of each
Doppler image into large-scale flows: differential rotation flows
$\VT_\mathrm{MF}$, meridional flows $\VT_\mathrm{MF}$, and the convective
blue-shift (limb shift) component $\VT_\mathrm{LS}$. This determines an
effective ``bias'' for each image according to
\begin{equation}
\VT_\mathrm{bias}=\VT_\mathrm{Rot}\left(B_0,\Phi,\Theta\right)+\VT_\mathrm{LS}'\left(\varrho\right)+\VT_\mathrm{MF}'\left(B_0,\Phi,\Theta\right),\label{eqn:V_bias}
\end{equation}
where $\Phi$ and $\Theta$ are the Stonyhurst coordinates
\cite[]{Thompson2006}, $B_0$ is the so-called ``Solar-B'' angle, and $\varrho$
is the heliocentric angle between the observer and the observed point (See
Appendix~\ref{sec:Helio}).  The form of~(\ref{eqn:eta})
and~(\ref{eqn:satellite}) suggests that artifacts introduced by the observer's
radial velocity will project onto the limb shift functions (cylindrical
symmetry), artifacts introduced by the observer's westward velocity will project
onto the differential rotation profile, and artifacts introduced by the
observer's northward velocity will project onto the meridional flow
profile. Because the geosynchronous orbit of \SDO{} causes significant
changes in the radial velocity $\VR$ over the period of a day, we
conjecture that the projection
of the radial velocity onto the limb shift functions is 
the most significant source of orbital artifacts in
Dopplergrams. Consequently, step 3 of the procedure consists of a
model for removing this effect, but first we must determine and
subtract the known biases from the each image. The following three
sections describe the standard mathematical formulation for capturing
the differential rotation, meridional flow, and the limb shift
``velocity.'' We then discuss, in detail, our application of this
formalism to the HMI data.  
\subsubsection{Rotational Velocity\label{sec:rotational}}
Following \cite{Hathaway1988} the solar differential rotational velocity is
decomposed by
\begin{equation}
U^0_\Phi\left(\Theta\right)=\sum_{\ell=1}^{\ell_\mathrm{max}}\,T_\ell^0\,\sqrt{\ell\,\left(\ell+1\right)}\,\Pbar_\ell^1\left(\sin\Theta\right)
\end{equation}
where
\begin{equation}
\Pbar_\ell^m\left(x\right)=\left(-1\right)^m\,\sqrt{\frac{\left(2\,\ell+1\right)\,\left(\ell-m\right)!}{2\,\left(\ell+m\right)!}}\,\Leg_\ell^m\left(x\right),\label{eqn:Pbar}
\end{equation}
and where $\Leg_\ell^m$ are the associated Legendre functions\footnote{as
  in Mathematica\textsuperscript{\textregistered} and Interactive Data Language (IDL)\textsuperscript{\textregistered}.}
\begin{equation}
\int\limits_{-1}^{1}{dx}\,\Leg_k^m\left(x\right)\,\Leg_\ell^m\left(x\right)=\frac{2\,\left(\ell+m\right)!}{\left(2\,\ell+1\right)\,\left(\ell-m\right)!}\,\delta_{k,\ell}.
\end{equation}
Note that~(\ref{eqn:Pbar}) leads to the definition of the spherical harmonics, 
\begin{equation}
Y_\ell^m\left(\theta,\phi\right)=\left(-1\right)^m\,\sqrt{\frac{\left(2\,\ell+1\right)}{4\,\pi}\,\frac{\left(\ell-m\right)!}{\left(\ell+m\right)!}}\,\Leg_\ell^m\left(\cos\theta\right)\,e^{i\,m\,\phi}\equiv\Pbar_\ell^m\left(\cos\theta\right)\,\frac{e^{i\,m\,\phi}}{{\sqrt{2\,\pi}}}.
\end{equation}
Using~(\ref{eqn:eta:los}), the projection of this solar differential rotation velocity onto the LOS
velocity is then
\begin{equation}
\VT_\mathrm{Rot}\left(B_0,\Phi,\Theta\right)=-\boldsymbol{\widehat{\eta}_\mathrm{LOS}}\cdot\Jacobian^\mathrm{T}\cdot\ephi\,U^0_\Phi\left(\Theta\right)
\end{equation}
where for $\theta_\rho\approx0$
\begin{equation}
\boldsymbol{\widehat{\eta}_\mathrm{LOS}}\cdot\Jacobian^\mathrm{T}\cdot\ephi\approx-\cos B_0\,\sin\Phi.
\end{equation}
 The coefficients $T_\ell^0$ are related to the
usual $A$, $B$, $C$ coefficients in
\begin{equation}
\omega\left(\Theta\right)=A+B\,\sin^2\Theta+C\,\sin^4\Theta=\frac{1}{R_\odot\,\cos\Theta}\,\sum_{\ell=1}^{\ell_\mathrm{max}}\,T_\ell^0\,\sqrt{\ell\,\left(\ell+1\right)}\,\Pbar_\ell^1\left(\sin\Theta\right)
\end{equation}
by
\begin{subequations}
\begin{align}
A&=\frac{1}{16} \left(8\,\sqrt{6}\,T_1^0-12\,\sqrt{14}\,T_3^0+15\,\sqrt{22}\,T_5^0\right),\label{eqn:rot_a}\\
B&= \frac{15}{8}\,\left(2\,\sqrt{14}\,T^0_3-7\,\sqrt{22}\,T^0_5\right),\\
C&=\frac{315}{8}\,\sqrt{\frac{11}{2}}\,T^0_5.
\end{align}
\end{subequations}
where $R_\odot=6.95946\time10^8$~km is the radius of the Sun for HMI
\cite[]{Emilio2015}.
\subsubsection{Limb Shift Velocity\label{sec:limbshift}}
Following \cite{Snodgrass1984} and \cite{Hathaway1992,Hathaway1996} the convective blue-shift,
caused by correlations between velocity and intensity in unresolved
convective elements \cite[]{Beckers1978}, is decomposed as
\begin{subequations}
\begin{equation}
\VT_\mathrm{LS}'\left(\varrho\right)=\sum_{\ell=0}^{N_{LS}}L_\ell\,\left(2\,\ell+1\right)^{-1/2}\,\mathcal{L}_\ell\left(1-\cos\varrho\right),\label{eqn:LS}
\end{equation}
where $\varrho$ is the heliocentric angle between the observer and the
observed point as defined in Appendix~\ref{sec:Helio}, and
\begin{equation}
\mathcal{L}_n\left(x\right)=\sqrt{2}\,\Pbar_\ell^0\left(2\,x-1\right),
\end{equation}
\end{subequations}
are shifted Legendre polynomials orthonormal on $\left(0,1\right)$. A prime is
used on~(\ref{eqn:LS}) to indicate that this is an adjusted limb shift function
for reasons that will become apparent below.
\subsubsection{Meridional Velocity\label{sec:meridional}}
Again, following \cite{Hathaway1992,Hathaway1996}, the meridional flows are decomposed as
\begin{equation}
U^0_\Theta\left(\Theta\right)=\sum_{\ell=1}^{\ell_\mathrm{max}}\,S_\ell^0\,\sqrt{\ell\,\left(\ell+1\right)}\,\Pbar_\ell^1\left(\sin\Theta\right),\label{eqn:MF}
\end{equation}
where positive coefficients correspond to northward velocities. In principle the
projection of the meridional flows is given by
\begin{equation}
\VT_\mathrm{MF}\left(B_0,\Phi,\Theta\right)=-\boldsymbol{\widehat{\eta}_\mathrm{LOS}}\cdot\Jacobian^\mathrm{T}\cdot\etheta\,U^0_\Theta\left(\Theta\right).
\end{equation}
However \cite{Hathaway1988} noticed that the meridional flow given
by~(\ref{eqn:MF}) has a projection onto the limb shift eigenfunctions. Thus,
to orthogonalize the description of the large scale flows, this projection
must be subtracted from the meridional flow eigenfunctions as it is already
been accounted for in the limb shift eigenfunctions. Thus, the actual
meridional flow description used to fit each image at stage~1 is given by
\begin{equation}
\VT_\mathrm{MF}'\left(B_0,\Phi,\Theta\right)=-\sum_{\ell=1}^{\ell_\mathrm{max}}\,S_\ell^0\,\sqrt{\ell\,\left(\ell+1\right)}\,\left[\boldsymbol{\widehat{\eta}_\mathrm{LOS}}\cdot\Jacobian^\mathrm{T}\cdot\etheta\,\Pbar_\ell^1\left(\sin\Theta\right)-G_\mathrm{MF,\ell}\left(B_0,\varrho\right)\right],\label{eqn:MF_FIT}
\end{equation}
where
\begin{equation}
G_{\mathrm{MF},\ell}\left(B_0,\varrho\right)=\frac{1}{2\,\pi}\,\int_0^{2\,\pi}d\psi\,\boldsymbol{\widehat{\eta}_\mathrm{LOS}}\cdot\Jacobian^\mathrm{T}\cdot\etheta\,\Pbar_\ell^1\left(\sin\Theta\right),\tag{\ref{eqn:GMF}}
\end{equation}
represents the spatial part of the meridional eigenfunction that is
independent of position angle $\psi$. Indeed,~(\ref{eqn:GMF}) is proportional
to the meridional eigenfunction averaged over the position angle $\psi$. The integral in~(\ref{eqn:GMF}) requires some effort to evaluate. The results
were stated first by \cite{Hathaway1988} without proof. A general proof in
closed form is presented in Appendix~\ref{sec:GMF}.
The eigenfunctions describing $\VT_\mathrm{LS}'$ are nearly orthogonal to
$\VT_\mathrm{MF}'$.\footnote{They are not exactly orthogonal because the
  observations on the disk are discrete and for simplicity we have
  orthogonalized these functions using continuous representations. See
  discussion in Section~\ref{sec:treatment} and Figure~\ref{fig:correlation}.}
Once the coefficients $L_m$ and $S_\ell^0$ are determined, the meridional flow
may be reconstructed with~(\ref{eqn:MF}) and the limb shift function may be
reconstructed with
\begin{equation}
\VT_\mathrm{LS}\left(B_0,\varrho\right)=\sum_{\ell=0}^{N_{LS}}\left[L_\ell\,\left(2\,\ell+1\right)^{-1/2}\,\mathcal{L}_\ell\left(1-\cos\varrho\right)+S_\ell^0\,\sqrt{\ell\,\left(\ell+1\right)}\,G_\mathrm{MF,\ell}\left(B_0,\varrho\right)\right].
\end{equation}
Note that 
\begin{equation}
\VT_\mathrm{LS}\left(B_0,\varrho\right)+\VT_\mathrm{MF}\left(B_0,\Phi,\Theta\right)=\VT_\mathrm{LS}'\left(B_0,\varrho\right)+\VT_\mathrm{MF}'\left(B_0,\Phi,\Theta\right).\label{eqn:equality}
\end{equation}
\subsubsection{Determination of the Bias\label{sec:bias}}
To determine the bias, each image is fit using the eigenfunctions described in
Sections~\ref{sec:rotational}\--\ref{sec:meridional}.
\cite{Hathaway1988,Hathaway1992,Hathaway1996} noted that strong magnetic
fields can alter the convection pattern. He used an iterative procedure that
first determines an estimate of the spectral coefficients and then replaces
the Doppler estimates in pixels corresponding to strong magnetic fields with
estimates consistent with these coefficients. The procedure repeats until
there are no further significant changes in the spectral coefficients.\par
The present analysis diverges significantly from that previous work. The image
at each stage of analysis is reformed into a column vectors
$\VLOS{1},\ldots,\VLOS{3}$ representing all of the $\ND$ pixels located on the
solar disk and $\vLOS{1},\ldots,\vLOS{3}$ representing just the $\NP$ weak
field pixels corresponding to an absolute LOS magnetic field less than or
equal to 10~G \textit{or} minimal support $\left(<60\right)$ as determined by
the disambiguation module ME\_720s\_fd10\{disambig\} of the vector
pipeline.\footnote{Strong field pixels correspond to absolute LOS magnetic
  field greater than 10~G \textit{or} significant support
  $\left(\ge60\right)$.} For the 17 days under consideration $\NP$ represents
about 84\% of the solar disk.  In the same manner, two matrices of
eigenfunctions are constructed $\EIG$ and $\eig$ representing the
$M=\NR+\NLS+\NMF$ eigenfunctions used to determine the large scale Doppler
patterns where $\EIG$ represents all of the $\ND$ pixels located on the solar
disk and $\eig$ represents just the $\NP$ weak field pixels. The goal is to
determine a set of spectral coefficients $\SC$ from the $\NP$ stage~1 Doppler
estimates in column vector $\vLOS{1}$.  Obviously with $\NP\simeq10^7$ and
$M\simeq24$ there is no unique solution to the overdetermined system
\begin{equation}
\begin{aligned}
\eig\SC=&\vLOS{1},\\
\left[\begin{matrix}\ee_{11}&\cdots&\ee_{1M}\\
\vdots&\cdots&\vdots\\
\ee_{\NP1}&\cdots&\ee_{\NP{M}}
\end{matrix}\right]\left[\begin{matrix}T_1^0\\ \vdots\\ T_{N_R}^0\\ L_0\\ \vdots\\ L_{N_\mathrm{LS}-1}\\ S_1^0\\ \vdots\\ S_{\NMF}^0\\ \end{matrix}\right]=&\left[\begin{matrix}v_{\mathrm{LOS\mhyphen1},1}\\ v_{\mathrm{LOS\mhyphen1},2}\\ v_{\mathrm{LOS\mhyphen1},3}\\ \vdots\\ v_{\mathrm{LOS\mhyphen1},\NP}\end{matrix}\right].\label{eqn:bias}
\end{aligned}
\end{equation}
where $v_{\mathrm{LOS\mhyphen1},i}$ are the $i=1,\ldots,\NP$ stage~1 weak
field pixels determined from~(\ref{eqn:satellite}) because no general inverse
of the eigenvectors $\eig$ exists. Instead we attempt to find the solution
which is best in a $L_2$ norm sense where $\SC^*$ minimizes
\begin{equation}
\left\Vert\eig\SC^*-\vLOS{1}\right\Vert_2\le\left\Vert\eig\SC-\vLOS{1}\right\Vert_2
\end{equation}
among all possible spectral coefficients $\SC$. This optimal solution is
determined from directly the least-squares solution
\begin{equation}
\SC^*=\left(\eig^\mathrm{T}\eig\right)^{-1}\eig^\mathrm{T}\vLOS{1},\label{eqn:LeastSquares}
\end{equation}
where $\left(\eig^\mathrm{T}\eig\right)^{-1}\eig^\mathrm{T}$ is known as the
pseudo-inverse \cite[]{Moore1920,Bjerhammar1951,Penrose1955}.  Note that this approach explicitly ignores the strong field
pixels in determining the spectral coefficients circumventing the iterations
necessary in \cite{Hathaway1996}. \par
The temporally varying bias image for each full-disk Dopplergram of $\ND$
pixels can then be reconstructed from
\begin{equation}
\bias=\EIG\SC^*.
\end{equation}
The large-scale bias free stage~2 Dopplergrams are then determined from
\begin{equation}
\VLOS{2}=\VLOS{1}-\EIG\SC^*
\end{equation}
\begin{figure}
\centerline{\includegraphics[width=4in]{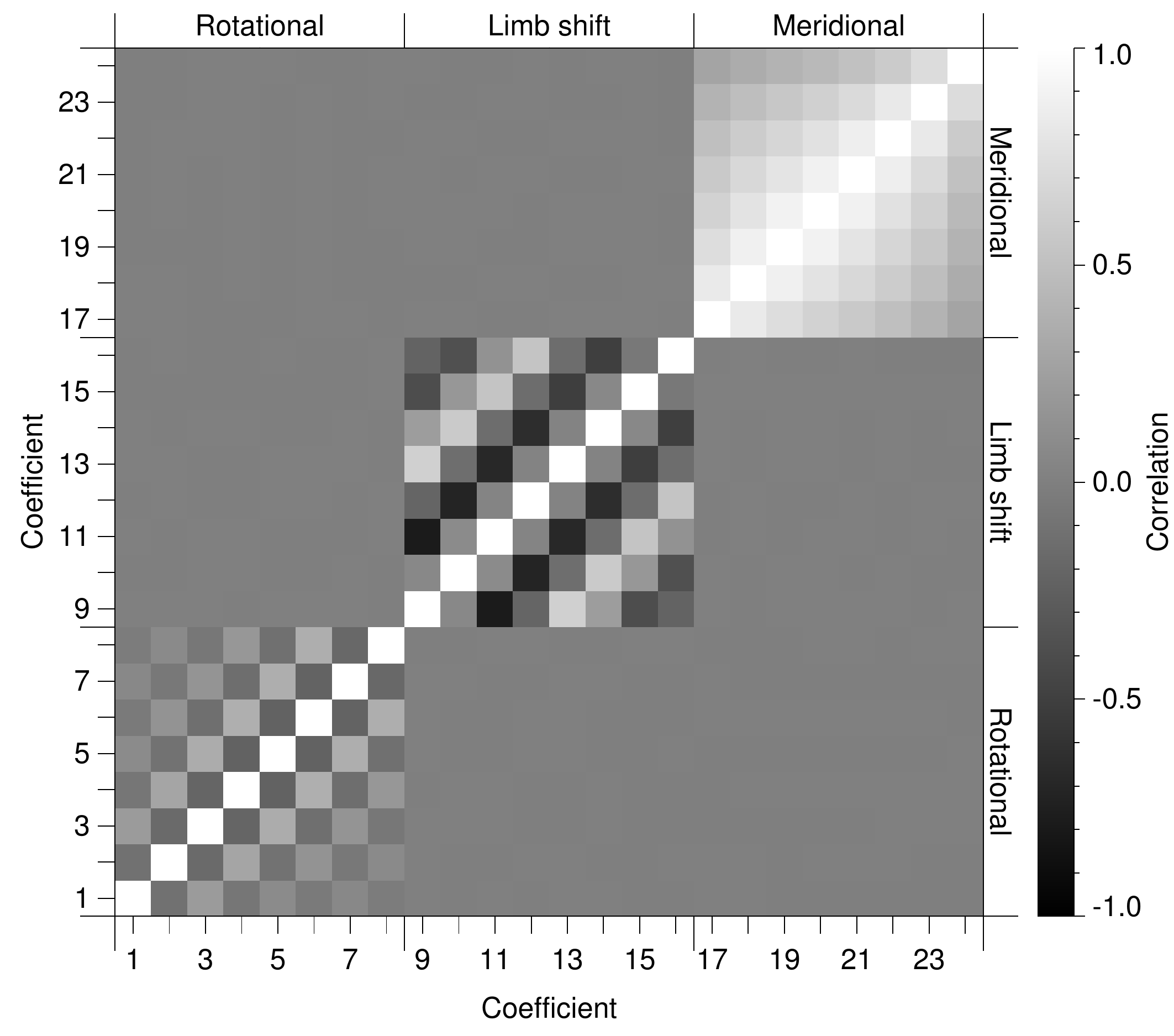}}
\caption{Correlations between the coefficients for the fit to the large scale
  flows on 2010-07-04 at 17:00:00. \label{fig:correlation}}
\end{figure}
Figure~\ref{fig:correlation} shows the correlations between the $M=24$
coefficients used to fit the large scale Doppler patterns. The eigenfunctions
are nearly block orthogonalized, \textit{e.g.} there are correlations between
the meridional eigenfunctions, but these eigenfunctions are completely
decoupled from the limb-shift eigenfunctions validating the corrections
encompassed by~(\ref{eqn:MF_FIT})\--(\ref{eqn:equality}). Correlations within
the block are expected as spherical harmonics themselves are not orthogonal on
the observed solar hemisphere and they are further confused by their
projection onto the LOS \cite[]{Mochizuki1992}.\par
\newlength{\coef}
\setlength{\coef}{5in}
\begin{figure}[!p]
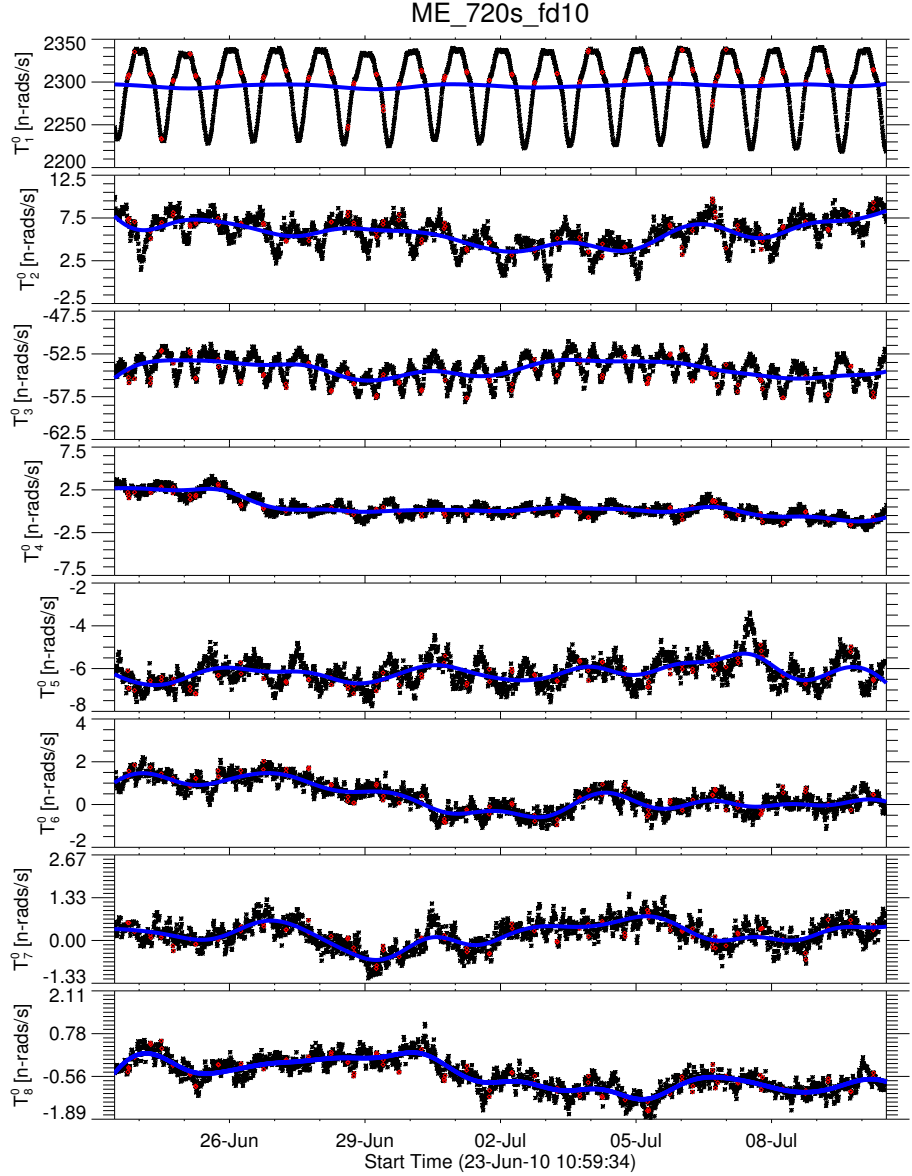

\centerline{\includegraphics[width=\coef]{{{f4}}}}
\caption{Rotational spectral coefficients during 17 days in 2010. The black
  data are ``high quality'' data and the red data are low quality data
  $\left(\mbox{Keyword: QUALITY}\neq0\right)$. The blue line corresponds to
  the low frequency trend determined after orbital artifacts are removed. See
  text for a complete discussion.\label{fig:rotational1}}
\end{figure}
\begin{figure}[!p]
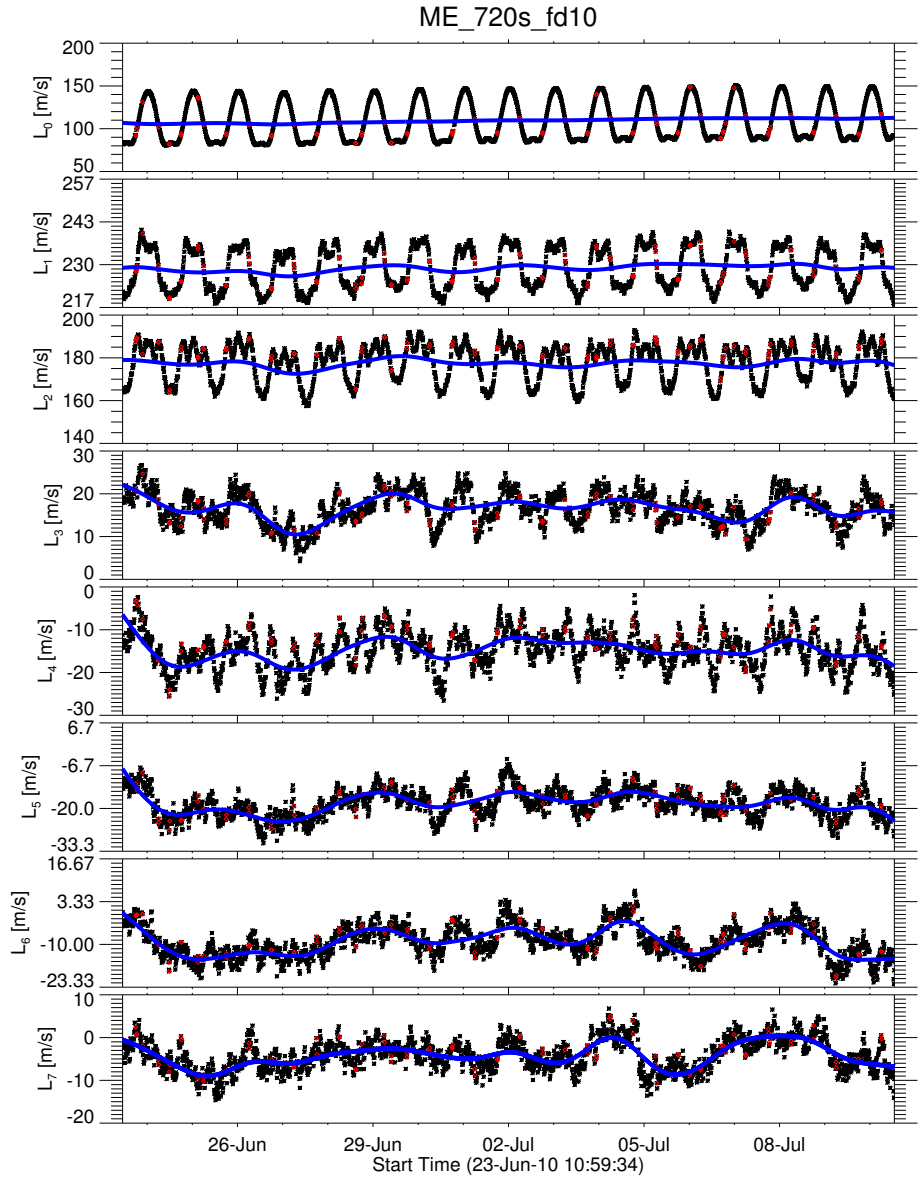

\centerline{\includegraphics[width=\coef]{{{f5}}}}
\caption{Limb shift spectral coefficients during 17 days in 2010. Same format at Figure~\ref{fig:rotational1}.\label{fig:limbshift1}}
\end{figure}
\begin{figure}[!p]
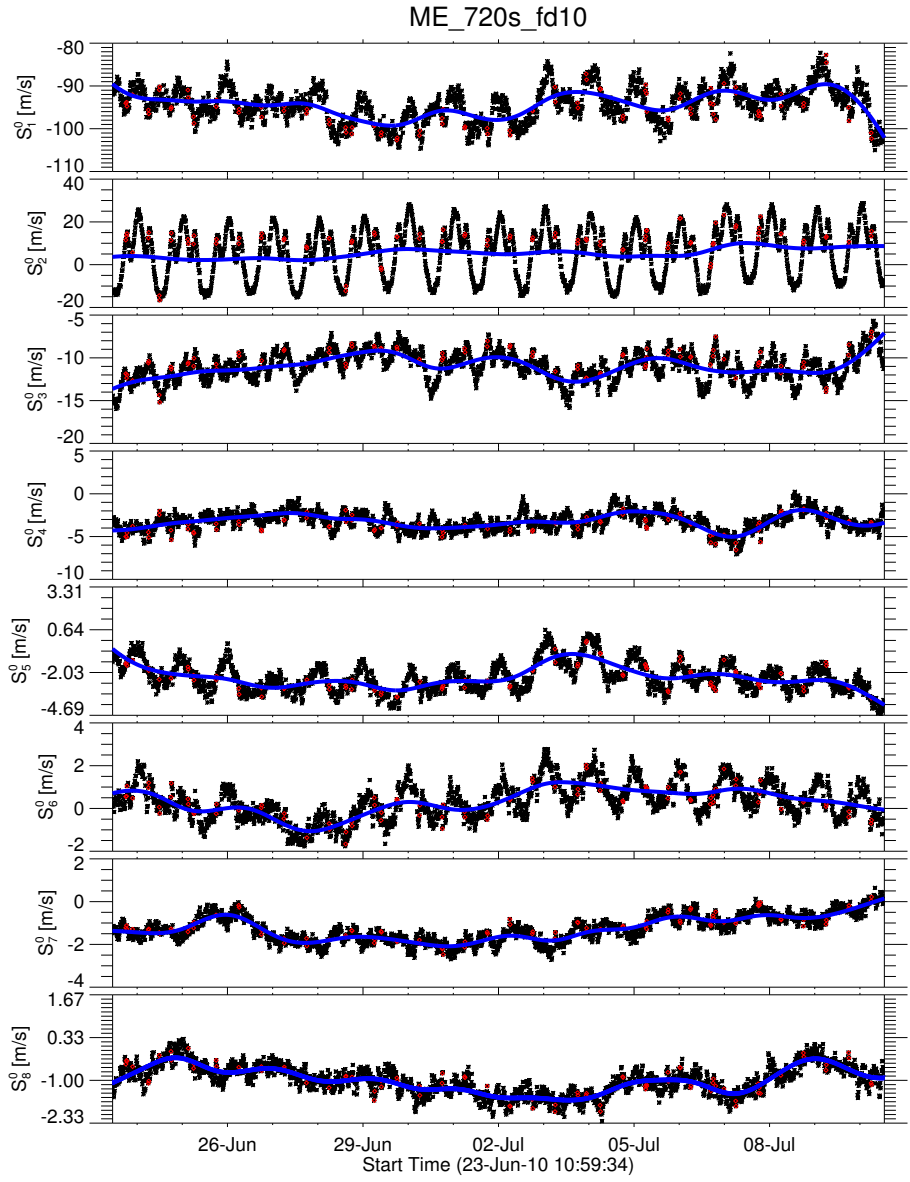

\centerline{\includegraphics[width=\coef]{{{f6}}}}
\caption{Meridional spectral coefficients during 17 days in 2010. Same format at Figure~\ref{fig:rotational1}.\label{fig:meridional1}}
\end{figure}
Figures~\ref{fig:rotational1}\--\ref{fig:meridional1} show rotational, limb
shift, and meridional spectral coefficients during 17 days in 2010 as
determined from~(\ref{eqn:LeastSquares}). The black data are ``high quality''
data and the red data are ``low quality'' data $\left(\mbox{Keyword:
  QUALITY}\neq0\right)$.  We emphasize that since the satellite velocity has
been removed from each pixel prior to fitting the Doppler data there should be
\textit{no correlation} with the satellite velocity. However, clear aperiodic
oscillations are present with a primary period of 24 hrs is observed in the
lowest spectral coefficients in all three Figures. In
Figure~\ref{fig:rotational1}, the lowest coefficient exhibits a peak-to-peak
amplitude $110\,$nrads/s or
$\Delta{v}=\sqrt{2}\,\Delta{T}_1^0\,R_\odot\simeq105$~m/s. Again, this
indicates a significant error in the measurements particularly near the limb
where solar rotation is the strongest in the LOS component.  The blue line
corresponds to the low-frequency trend with a cutoff period of 48~hrs
determined from a nonparametric B-spline filter
\cite[]{Schuck2010,Woltring1986}. The details of how this trend is determined
is discussed in Section~\ref{sec:treatment}. \par
The red curve in Figure~\ref{fig:raw_doppler} shows the residual spatially
averaged RMS Doppler velocity for the stage~2 Dopplergrams
$\VLOS{2}$. Removing the time-vary bias considerably improves the temporal
stability of the Dopplergrams.  One of the most striking features of these
data is the bias in the $S_1^0$ meridional coefficient exhibited by
Figures~\ref{fig:meridional1}. One explanation for this result is an
average cross-equatorial flow of $\simeq-95$~m/s. A cross-equatorial flow of
this magnitude would be easily detected by other techniques, so this
explanation is ruled out. Another possible explanation for this bias is an
error in position angle $\delta\psi=\tan^{-1}\left(S_1^0/A\right)$ in solar
North where $A$ is determined from~(\ref{eqn:rot_a}). A bias of
$\simeq-95$~m/s corresponds to error in position angle of about
$2.6^\circ$. This explanation also seems incorrect given the agreement between
the alignment of HMI and MDI magnetogram data exhibited in
Figure~\ref{fig:mdi_roll}. Therefore, our conclusion at present is that this
bias in the $S_1^0$ coefficient represents a spatial non-uniformity in the
response of HMI.\par
\subsection{Stage~3: Gain Adjustment}
\begin{figure}[!p]
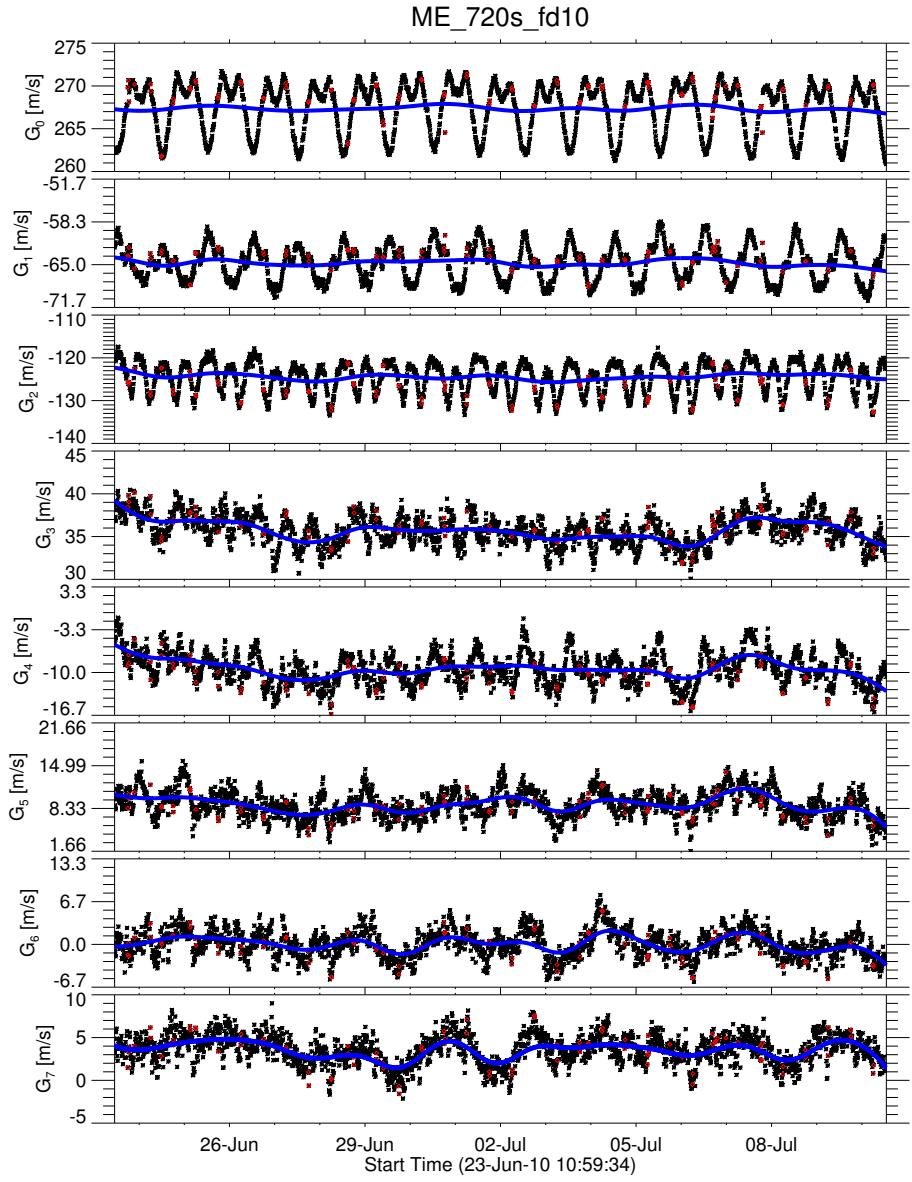

\centerline{\includegraphics[width=\coef]{{{f8}}}}
\caption{Gain coefficients during 17 days in 2010. Same format as in Figure~\ref{fig:rotational1}.\label{fig:gain1}}
\end{figure}
Figure 3 shows that even after the known velocities are removed from
the HMI data, there still remains a strong daily periodicity in the
residual images, indicating contamination by orbital
effects. Therefore, stage~3 of our \CODERED{} procedure consists
of a pixel by pixel, image by image, adjustment of the gain. Given the strong
aperiodic oscillations in the limb shift functions, we hypothesize
that the gain of each pixel follows a similar pattern; consequently,
we use the limb shift formalism to correct for this gain. Following the
procedure in Section~\ref{sec:bias}, two $\REAL{\ND}{\NGain}$ and
$\REAL{\NP}{\NGain}$ matrices are constructed from the limb shift
eigenfunctions,
$\EIG_\mathrm{LS}=\EIG\left(1:\ND,\NR+1:\NR+\NLS+1\right)$ and
$\eig_\mathrm{LS}=\eig\left(1:\NP,\NR+1:\NR+\NLS+1\right)$,
corresponding to the $\NGain=\NLS$ limb shift eigenfunctions used to
fit the bias where $\EIG_\mathrm{LS}$ represents all of the $\ND$
pixels located on the solar disk and $\eig_\mathrm{LS}$ represents
just the $\NP$ weak field pixels. These eigenfunctions are then fit to
the absolute value of the stage~2 bias subtracted Dopplergrams
\begin{equation}
\begin{aligned}
\eig_\mathrm{LS}\GC=&\left|\vLOS{2}\right|,\\
\left[\begin{matrix}\ee_{10}&\cdots&\ee_{1\NGain-1}\\
\vdots&\cdots&\vdots\\
\ee_{\NP0}&\cdots&\ee_{\NP\NGain-1}
\end{matrix}\right]\left[\begin{matrix}G_0\\ \vdots\\ G_{N_\mathrm{G}-1}\end{matrix}\right]=&\left[\begin{matrix}
\abs{v_{\mathrm{LOS\mhyphen2},1}}\\ 
\abs{v_{\mathrm{LOS\mhyphen2},2}}\\ \vdots\\ 
\abs{v_{\mathrm{LOS\mhyphen2},\NP}}
\end{matrix}\right],\label{eqn:gain}
\end{aligned}
\end{equation}
which has the optimal least-squares solution
\begin{equation}
\GC^*=\left(\eig_\mathrm{LS}^\mathrm{T}\eig_\mathrm{LS}\right)^{-1}\eig_\mathrm{LS}^\mathrm{T}\left|\vLOS{2}\right|.\label{eqn:LeastSquares:G}
\end{equation}
The gain for each image can then be reconstructed from
\begin{equation}
\gain=\EIG_\mathrm{LS}\GC^*.
\end{equation}
Figure~\ref{fig:gain1} shows the gain coefficients during 17 days in 2010 as
determined from~(\ref{eqn:LeastSquares:G})
\subsection{Treatment of the Coefficients and Image Reconstruction\label{sec:treatment}}
\begin{figure}[!p]
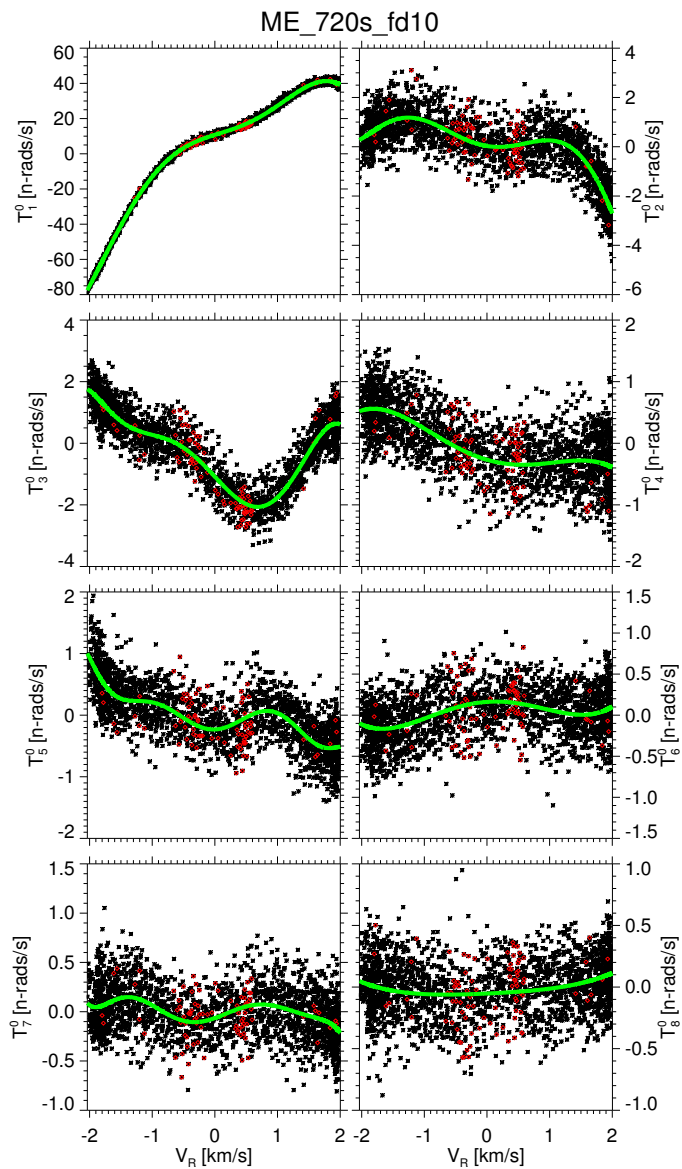

\centerline{\includegraphics[width=\coef]{{{f9}}}}
\caption{Rotational spectral coefficients during 17 days in 2010 as a function of
  satellite radial velocity $\VR$. The black data are ``high quality'' data
  and the red data are low quality data $\left(\mbox{Keyword:
    QUALITY}\neq0\right)$. The green line corresponds to the ``best'' fit of
  orthogonalized polynomials as determined by the BIC.\label{fig:rotational2}}
\end{figure}
\begin{figure}[!p]
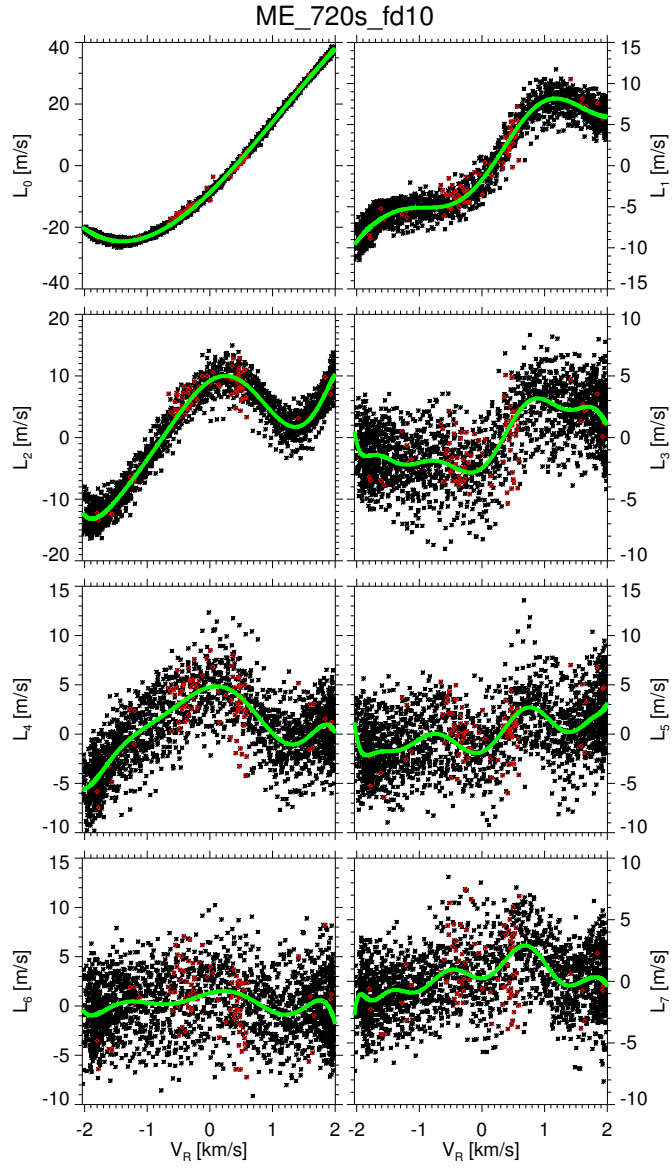

\centerline{\includegraphics[width=\coef]{{{f10}}}}
\caption{Limb shift spectral coefficients during 17 days in 2010 as a function of
  satellite radial velocity $\VR$. Same format at
  Figure~\ref{fig:rotational2}.  \label{fig:limbshift2}}
\end{figure}
\begin{figure}[!p]
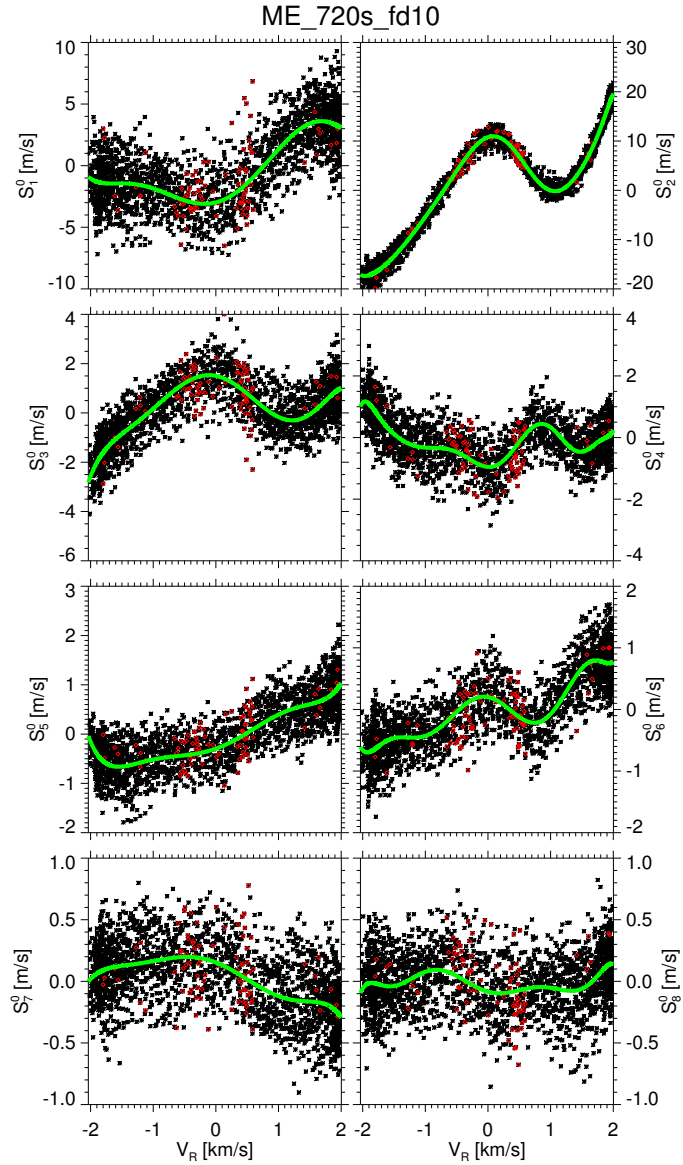

\centerline{\includegraphics[width=\coef]{{{f11}}}}
\caption{Meridional spectral coefficients during 17 days in 2010 as a function of
  satellite radial velocity $\VR$. Same format at
  Figure~\ref{fig:rotational2}. \label{fig:meridional2}}
\end{figure}
\begin{figure}[!p]
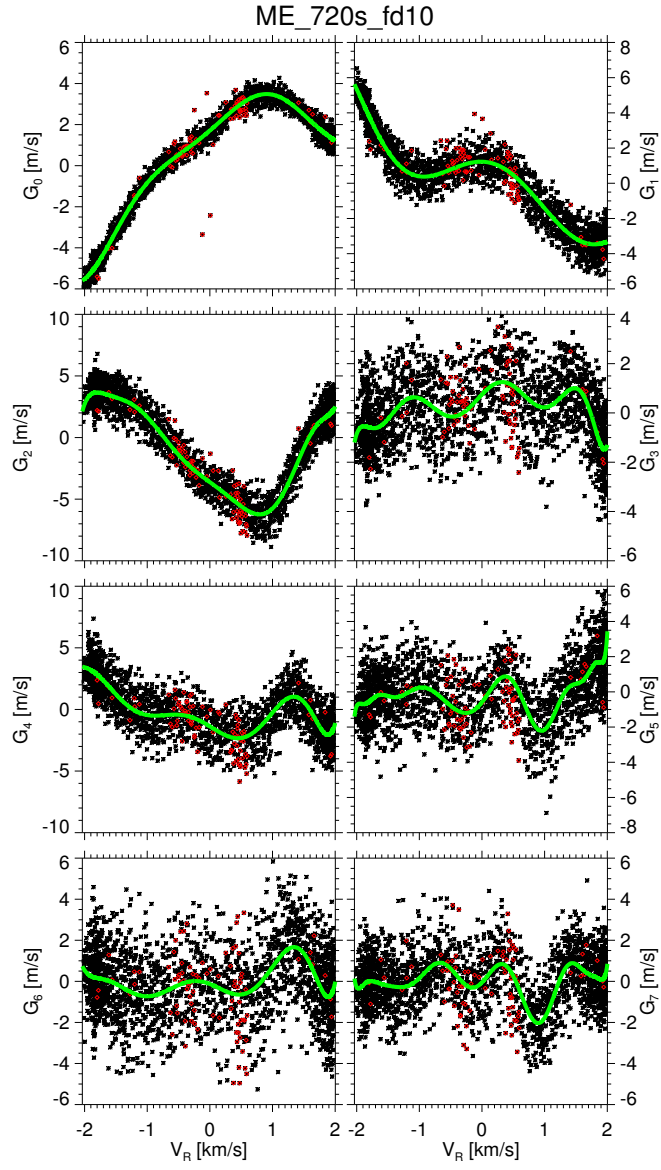

\centerline{\includegraphics[width=\coef]{{{f12}}}}
\caption{Gain coefficients during 17 days in 2010 as a function of
  satellite radial velocity $\VR$. Same format at
  Figure~\ref{fig:rotational2}. \label{fig:gain2}}
\end{figure}
Plotting the data in Figures~\ref{fig:rotational1}\--\ref{fig:gain1} as a
function of $\VR$, reveals their systematic \textit{nonlinear} dependence on
radial satellite velocity as shown in
Figures~\ref{fig:rotational2}\--\ref{fig:gain2}. The nonlinear response is the
source of the aperiodic temporal dependence in the coefficients. The black
data points in these figures are fit by weighted least squares with
orthogonalized polynomials.\footnote{The red ``low quality'' data are
  ignored.} The weights for each data point are determined from the variance
in the coefficients from the fits. The optimal polynomial order is determined
using the Bayesian Information Criteria \cite[]{Schwarz1978,Ming2008}
\begin{equation}
\mathrm{BIC}_k\simeq
N_\mathrm{D}\,\log\widehat\sigma^2_\mathrm{ML}+N_k\,\log{N_\mathrm{D}},\label{eqn:BIC}
\end{equation}
where $N_\mathrm{D}$ is the number of data, $N_k$ is the number of model
parameters and $\widehat\sigma^2_\mathrm{ML}$ is the maximum likelihood
estimate of the variance. The optimal number of coefficients $k_\mathrm{opt}$
corresponds to the model with the minimum $\mathrm{BIC}_k$ value.\footnote{The
  application of BIC does not require that the true model is in the set of
  models (polynomials) under consideration \cite[]{Cavanaugh1999}.} We
emphasize that the ``model'' involves two separate fitting processes: (1) The
orthogonalized polynomial fit to the parameterized data in $V_R$ space and (2)
the temporal fit using a nonparametric B-spline filter to remove any
systematic temporal drift in the parameters which would bias the
orthogonalized polynomial fit. The B-spline filter has some attractive
properties for this problem as data frames are missing and a B-spline filter
can be interpreted as an optimal cascaded Butterworth filter generalized for
unevenly sampled data \cite[]{Craven1979}. The number of degrees of freedom
removed from the data by the smoothing procedure can be determined from the
trace of the influence matrix of the B-spline filter
\cite[]{Wahba1980,Woltring1986}. The maximum likelihood estimate of the error
variance is computed from the residuals by subtracting both of these fits from
the stage~1 data and therefore $N_k$ in~(\ref{eqn:BIC}) must reflect both of
these fitting processes, \textit{i.e.}  the order of the orthogonalized
polynomial and the trace of the influence matrix of the B-spline filter. The
smoothing parameter of the B-spline filter is fixed with a cutoff of 48~hrs
and some iteration is necessary to minimize the BIC and find the optimal
coefficients for each polynomial order $k$.\par
Using the low frequency response and the orthogonalized polynomial fit
corresponding to the blue and green curves in
Figures~\ref{fig:rotational1}\--\ref{fig:gain1}, a model for the
\textit{predicted} bias and gain coefficients may be determined for any radial
velocity $\VR$ time $t$ in the data set
\begin{subequations}
\begin{align}
\BIAS\left(\VR,t\right)&=\BIAS_\mathrm{LF}\left(t\right)+\BIAS_\mathrm{OP}\left(\VR\right),\\
\GAIN\left(\VR,t\right)&=\underbrace{\GAIN_\mathrm{LF}\left(t\right)}_\mathrm{blue}+\underbrace{\GAIN_\mathrm{OP}\left(\VR\right)}_\mathrm{green}.
\end{align}
The stage~3 Dopplergrams are reconstructed by subtracting off the
\textit{observed} bias and dividing by the \textit{observed} gain, then
multiplying by the \textit{predicted} gain and adding back the
\textit{predicted} bias at a consistent radial velocity $\VR\equiv0$
\begin{equation}
\VLOS{3}=\left[\EIG_\mathrm{LS}\GAIN\left(0,t\right)\right]\oslash\left[\EIG_\mathrm{LS}\GC^*\right]\odot\left[\VLOS{1}-\EIG\SC^*\right]+\EIG\BIAS\left(0,t\right),\label{eqn:VLOS3}
\end{equation}
\end{subequations}
where $\oslash$ and $\odot$ represent Hadamard (element-wise) division and
multiplication of vectors. \par
These stage~3 Dopplergrams are the final result of our \CODERED{}
procedure. It is important to note that ~(\ref{eqn:VLOS3}) does not
\textit{correct} the images in some absolute sense; however, it does
remove the \textit{inconsistencies} caused by the orbital artifacts. The blue
curve in Figure~\ref{fig:raw_doppler} shows the residual spatially averaged
RMS Doppler velocity for the stage~3 Dopplergrams $\VLOS{2}$. Removing the
time-vary bias and spatially and temporally adjusting the gain greatly
improves the temporal stability of the Dopplergrams. The difference in
variability between the black curve (stage 1) and blue curve (stage 3) is
roughly 31dB.\par
The reduction in RMS velocity shown in Figure~\ref{fig:raw_doppler} is
striking. This result indicates that the \CODERED{} procedure is
highly effective at removing orbital effects, at least, on large
scales. The key questions, however, are whether it is robust in that
it removes artifacts from all scales and whether it is accurate in
that the procedure introduces no new artifacts to the data. In order
to answer these question, we perform below a detailed spectral analysis
of the images after each \CODERED{} stage. This analysis shows that
the procedure does clean the data from orbital artifacts at all spatial scales
and that it does not contaminate the data with new artifacts.

\makeatletter{}\section{Analysis of Stages\--1\---3}
\begin{figure}[!p]
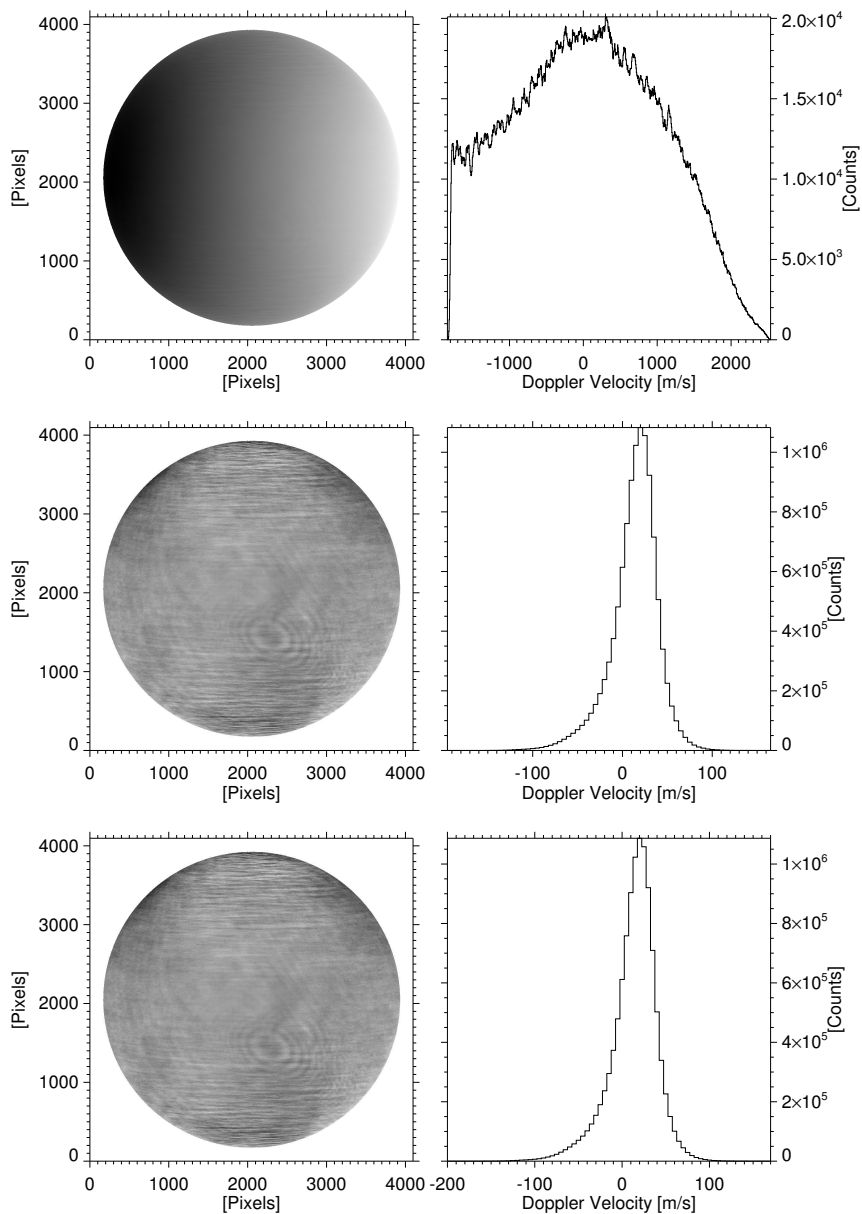

\centerline{\includegraphics[width=4.5in]{{{f13a}}}}
\centerline{\includegraphics[width=4.5in]{{{f13b}}}}
\centerline{\includegraphics[width=4.5in]{{{f13c}}}}
\caption{Median Co-aligned Doppler image and histogram after various stages of
  processing: The top is stage 1, the middle is stage 2, and the bottom is
  stage 3. The stage 1 image is dominated by the time-averaged differential
  rotation, limb shift and meridional flow patterns. The middle and bottom
  panels exhibit a clear Fresnel pattern.\label{sec:median}}
\end{figure}
\SDO{} produces an overwhelming amount of data. Even for the reduced data
series considered here, there are $20\times10^{9}$ pixels to analyze over the
17 day period.  Further complicating the analysis, data frames are lost in any
long data series from \SDO{}, and consequently the sample rates are
nonuniform. These data require techniques that scale well with
$\ND\simeq10^7$, the number of pixels in each image and $\Nt\simeq10^3$, the
number of images and methods that are either insensitive to the sample rates
or that are explicitly designed to analyze non-uniformly sampled data. For
this data set we implement the Karhunen-Lo\'eve (KL) transform
\cite[]{Loeve1955,Lumley1967,Sirovich1987,Holmes1996} briefly described in
Appendix~\ref{sec:KL} combined with the CLEAN algorithm for computing power
spectral density of unevenly sampled time-series
\cite[]{Roberts1987,Hogbom1974}. The KL transform decomposes the dynamics into
a set of $\ND\times\Nt$ orthogonal spatial modes $\PHI$ and $\Nt\times\Nt$
temporal coefficients $\ALPHA$ with $\Nt\ll\ND$. The advantage of this
analysis is that the temporal dynamics of the entire image sequence is
represented by a relatively small matrix $\ALPHA$ in contrast to attempting to
interpret to spectral properties of $\ND\times\Nt$ pixels in the image
sequence. The disadvantages of the technique are that the spatial information
associated with the dynamics is decoupled from the temporal dynamics and the
spatial modes are purely empirical \--- there isn't always an simple physical
interpretation for the spatial structure of the modes. Using the CLEAN
algorithm on the coefficient matrix $\ALPHA$ completely characterizes the
spectral properties of the data as the spatial eigenfunctions $\PHI$ are
orthogonal.\par
The images for the various stages are first co-aligned and registered into a
data cube or image sequence denoted $\IS'$ which is $\REAL{\ND}{\Nt}$. The
median is determined for each stage via
\begin{subequations}
\begin{equation}
\left\langle\IS\right\rangle_i=\median_j\left(\IS_{ij}'\right),
\end{equation}
where $\left\langle\IS\right\rangle_i$ is understood to be the temporal median
of each pixel in $\Nt$ images.  This median is then subtracted from each image
\begin{equation}
\IS_{ij}=\IS'_{ij}-\left\langle\IS\right\rangle_i.
\end{equation}
\end{subequations}
Figure~\ref{sec:median} shows the median co-aligned Doppler image and histogram
after various stages of processing: The top is stage 1, the middle is stage 2,
and the bottom is stage 3. The stage 1 image is dominated by the time-averaged
differential rotation, limb shift and meridional flow patterns. The middle and
bottom panels exhibit a clear Fresnel pattern.\par
\begin{figure}
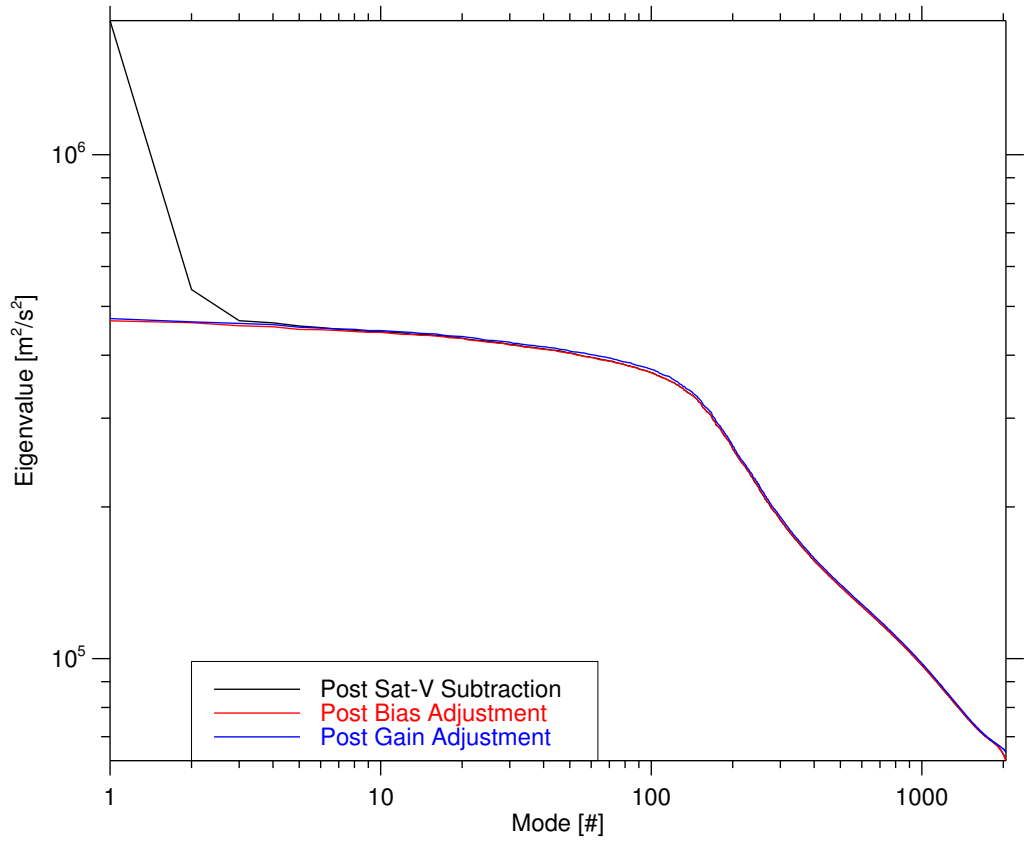

\centerline{\includegraphics[width=6in,clip=]{{{f14}}}}
\caption{KL eigenvalue spectra after
the various stages of analysis: 1 (black), 2 (red) and 3 (blue).\label{fig:KL_EIGVAL}}
\end{figure}
The KL transform is used to decompose this image sequence into a
$\REAL{\Nt}{\Nt}$ matrix of orthogonal coefficients $\ALPHA$ and an
$\REAL{\ND}{\Nt}$ matrix of orthogonal spatial modes $\PHI$. In general, the
KL modes have the following temporal and spatial properties $\Delta
T\simeq1/\#$ and $\Delta L\simeq1/\#$, \textit{i.e.}, higher mode numbers
(\#'s) correspond to faster time-scales and smaller
spatial-scales. Figure~\ref{fig:KL_EIGVAL} shows the eigenvalue spectra after
the various stages of analysis: 1 (black), 2 (red), and 3 (blue). While stages
2 and 3 are very similar, stage 1 which contains both orbital effects in the
large scale bias and the small scale velocities is significantly different for
the three lowest modes and perhaps as high as mode \#5. This eigenvalue
spectra can be interpreted as ``variance explained'' by each of the modes.
\begin{figure}[t]
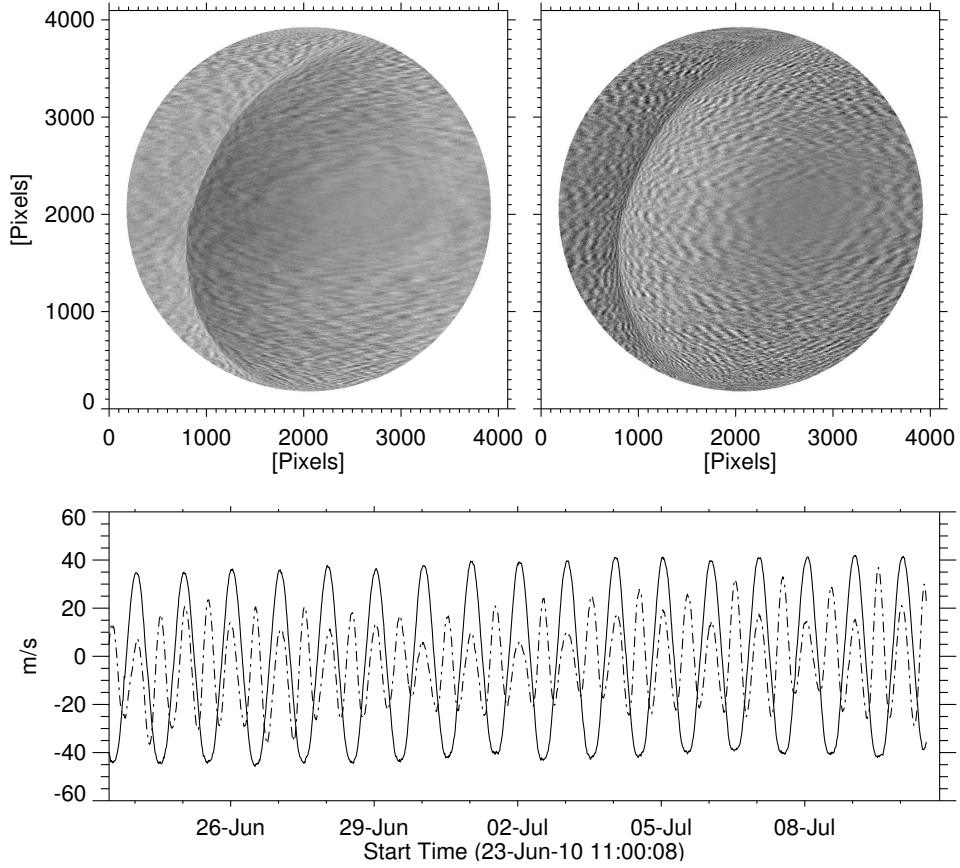

\centerline{\includegraphics[width=5in,viewport=16 164 523 622,clip=]{{{f15}}}}
\caption{Top: spatial KL modes \#1 and \#2 for stage 1. Bottom: time history
  of the coefficients for modes \#1 (solid) and \#2 (dashed). Note the oddly
  shaped distribution of spatial power and the nearly periodic behavior of the
  coefficients.\label{fig:stage1_klmodes}}
\end{figure}
\begin{figure}[t]
\centerline{\includegraphics[width=5in,viewport=16 164 523 622,clip=]{{{f16}}}}
\caption{Top: spatial KL modes \#1 and \#2 for stage 2. Bottom: time history
  of the coefficients for modes \#1 (solid) and \#2 (dashed).\label{fig:stage2_klmodes}}
\end{figure}
\begin{figure}[t]
\centerline{\includegraphics[width=5in,viewport=16 164 523 622,clip=]{{{f17}}}}
\caption{Top: spatial KL modes \#1 and \#2 for stage 3. Bottom: time history
  of the coefficients for modes \#1 (solid) and \#2 (dashed).\label{fig:stage3_klmodes}}
\end{figure}
Figures~\ref{fig:stage1_klmodes}\--\ref{fig:stage3_klmodes} show the spatial
eigenfunctions and temporal behavior of the coefficients of modes \#1 and \#2
for stages 1, 2, and 3.  Note the oddly shaped distribution of spatial power
and the nearly periodic behavior of the coefficients cause by the time-varying
bias and gain in Figure~\ref{fig:stage1_klmodes} for stage 1.\par
\newlength{\width}
\newlength{\heightone}
\newlength{\heighttwo}\setlength{\heightone}{1.7996in}
\setlength{\heighttwo}{ 0.725451in}
\setlength{\width}{2.4in}
\setlength{\heightone}{2.20451in}
\setlength{\heighttwo}{0.888677in}
\newcommand*{\toplabel}{0pt 65pt 110pt 514pt}
\newcommand*{\topviewport}{110pt 65pt 609pt 514pt}
\newcommand*{\topcolorbar}{609pt 65pt 711pt 514pt}
\newcommand*{\middlelabel}{0pt 74pt 110pt 255pt}
\newcommand*{\middleviewport}{110pt 74pt 609pt 255pt}
\newcommand*{\middlecolorbar}{609pt 74pt 711pt 255pt}
\newcommand*{\bottomlabel}{110pt 13pt 609pt 74pt}
\begin{sidewaysfigure}
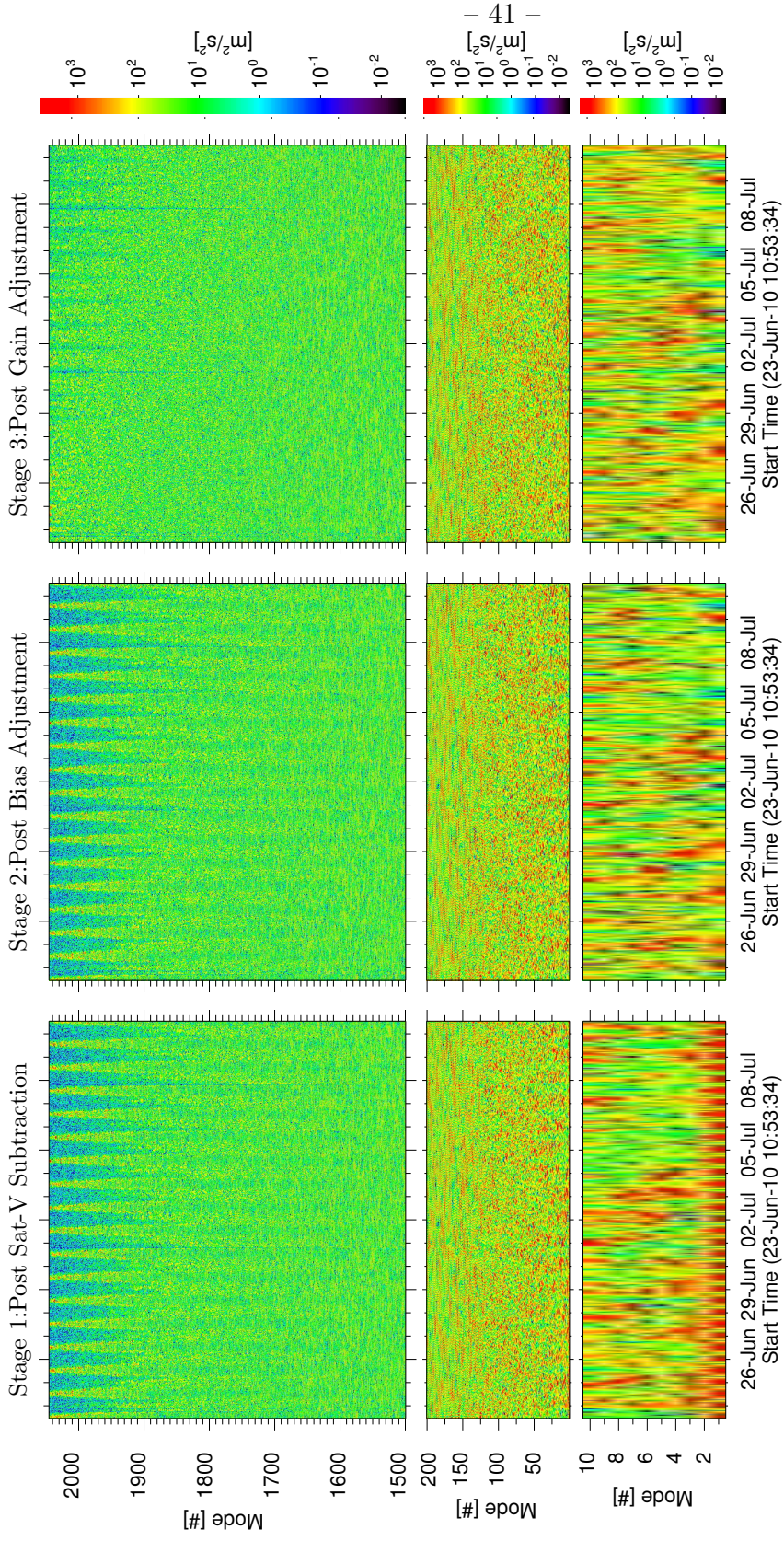

\setlength\tabcolsep{0pt}
\renewcommand{\arraystretch}{0} 
\begin{center}
\footnotesize
\begin{tabular}{ccccc}
&Stage 1:Post Sat-V Subtraction&Stage 2:Post Bias Adjustment&Stage 3:Post Gain Adjustment\\
\includegraphics[height=\heightone,clip=,viewport=\toplabel]{{{f18d}}}&
\includegraphics[width=\width,clip=,viewport=\topviewport]{{{f18d}}}&
\includegraphics[width=\width,clip=,viewport=\topviewport]{{{f18e}}}&
\includegraphics[width=\width,clip=,viewport=\topviewport]{{{f18f}}}&
\includegraphics[height=\heightone,clip=,viewport=\topcolorbar]{{{f18f}}}\\ 
\includegraphics[height=\heighttwo,clip=,viewport=\middlelabel]{{{f18g}}}&
\includegraphics[width=\width,clip=,viewport=\middleviewport]{{{f18g}}}&
\includegraphics[width=\width,clip=,viewport=\middleviewport]{{{f18h}}}&
\includegraphics[width=\width,clip=,viewport=\middleviewport]{{{f18i}}}&
\includegraphics[height=\heighttwo,clip=,viewport=\middlecolorbar]{{{f18i}}}\\
\includegraphics[height=\heighttwo,clip=,viewport=\middlelabel]{{{f18k}}}&
\includegraphics[width=\width,clip=,viewport=\middleviewport]{{{f18k}}}&
\includegraphics[width=\width,clip=,viewport=\middleviewport]{{{f18l}}}&
\includegraphics[width=\width,clip=,viewport=\middleviewport]{{{f18m}}}&
\includegraphics[height=\heighttwo,clip=,viewport=\middlecolorbar]{{{f18m}}}\\
&
\includegraphics[width=\width,clip=,viewport=\bottomlabel]{{{f18g}}}&
\includegraphics[width=\width,clip=,viewport=\bottomlabel]{{{f18h}}}&
\includegraphics[width=\width,clip=,viewport=\bottomlabel]{{{f18i}}}
\end{tabular}
\end{center}
\caption{Power in the coefficients of the KL transforms
  $\left(\ALPHA^2\right)$ for stage 1 (left column), stage 2 (middle column),
  and stage 3 (right column) data. The top/middle/bottom panels correspond to
  a magnified view of modes 1500\--2045, 1\--200, and 1\--10 for each stage.\label{fig:KL_ALPHA}}
\end{sidewaysfigure}
Figure~\ref{fig:KL_ALPHA} shows the power in the coefficients of the KL
transforms $\left(\ALPHA^2\right)$ as a function of time for the stage 1 (left
column), stage 2 (middle column), and stage 3 (right column) data.  The
top/middle/bottom panels correspond to a magnified view of modes 1500\--2045,
1\--200, and 1\--10 lowest modes for each stage. The stage 1 data
(left column) exhibits a clear daily periodicities as expected from
Figure~\ref{fig:stage1_klmodes} and an overall daily modulation of the power
particularly evident below mode 50 in the middle panel and above mode 1700 in
the top panel. The stage 2 data (middle column), after the large scale flows
have been removed, is improved at the lowest few modes in the bottom panel
reinforcing the concept that the lowest KL modes correspond to the slowest
time-scales and the largest spatial scales. The stage 3 data (right column),
after gain adjustment, is dramatically more uniform than the stage~1 and
stage~2 data in the top and middle panels. Despite the improvement, there
remain some temporal artifacts in the power of the highest mode numbers for
the stage 3 data.\par
\setlength{\width}{1.7in}
\renewcommand*{\toplabel}{0pt 36pt 110pt 514pt}
\renewcommand*{\topviewport}{110pt 36pt 609pt 514pt}
\renewcommand*{\topcolorbar}{611pt 36pt 707pt 514pt}
\renewcommand*{\heightone}{1.91583in}
\renewcommand*{\bottomlabel}{110pt 6pt 609pt 36pt}
\renewcommand*{\middlelabel}{0pt 43pt 110pt 255pt}
\renewcommand*{\middleviewport}{110pt 43pt 609pt 255pt}
\renewcommand*{\middlecolorbar}{611pt 43pt 707pt 255pt}
\renewcommand*{\heighttwo}{0.849699in}

\begin{figure}
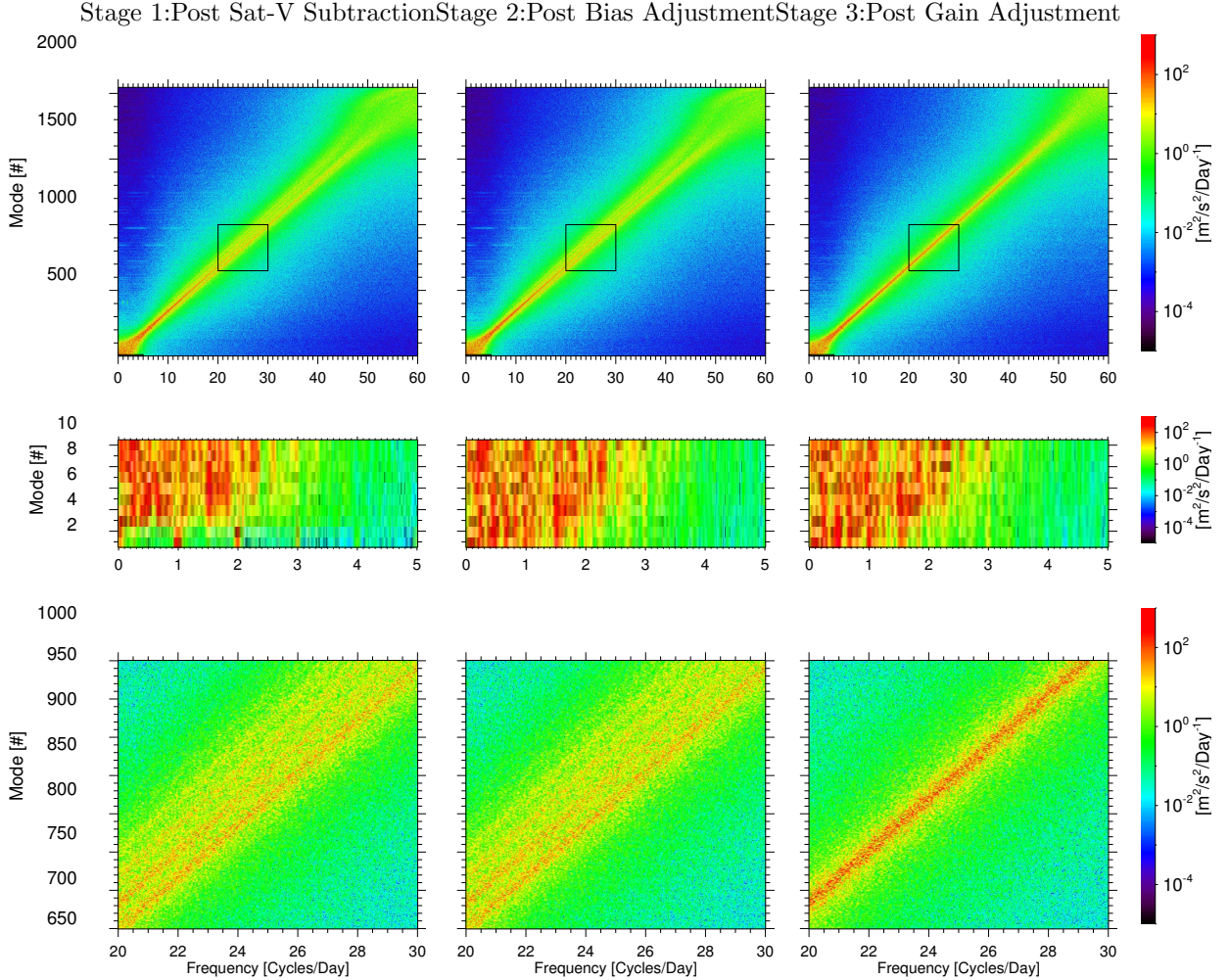

\setlength\tabcolsep{0pt}
\renewcommand{\arraystretch}{0} 
\begin{center}
\footnotesize
\begin{tabular}{ccccc}
&Stage 1:Post Sat-V Subtraction&Stage 2:Post Bias Adjustment&Stage 3:Post Gain Adjustment\\
\includegraphics[height=\heightone,clip=,viewport=\toplabel]{{{f19a}}}&
\includegraphics[width=\width,clip=,viewport=\topviewport]{{{f19a}}}&
\includegraphics[width=\width,clip=,viewport=\topviewport]{{{f19b}}}&
\includegraphics[width=\width,clip=,viewport=\topviewport]{{{f19c}}}&
\includegraphics[height=\heightone,clip=,viewport=\topcolorbar]{{{f19c}}}\\ \noalign{\vskip0.15in}
\includegraphics[height=\heighttwo,clip=,viewport=\middlelabel]{{{f19d}}}&
\includegraphics[width=\width,clip=,viewport=\middleviewport]{{{f19d}}}&
\includegraphics[width=\width,clip=,viewport=\middleviewport]{{{f19e}}}&
\includegraphics[width=\width,clip=,viewport=\middleviewport]{{{f19f}}}&
\includegraphics[height=\heighttwo,clip=,viewport=\middlecolorbar]{{{f19f}}}\\ \noalign{\vskip0.15in}
\includegraphics[height=\heightone,clip=,viewport=\toplabel]{{{f19g}}}&
\includegraphics[width=\width,clip=,viewport=\topviewport]{{{f19g}}}&
\includegraphics[width=\width,clip=,viewport=\topviewport]{{{f19h}}}
&
\includegraphics[width=\width,clip=,viewport=\topviewport]{{{f19i}}}&
\includegraphics[height=\heightone,clip=,viewport=\topcolorbar]{{{f19i}}}\\
&
\includegraphics[width=\width,clip=,viewport=\bottomlabel]{{{f19a}}}&
\includegraphics[width=\width,clip=,viewport=\bottomlabel]{{{f19b}}}&
\includegraphics[width=\width,clip=,viewport=\bottomlabel]{{{f19c}}}
\end{tabular}
\end{center}
\caption{Power spectral density in the coefficients of the KL transforms
  $\left(\ALPHA^2\right)$ for stage~1 (left column), stage~2 (middle column),
  and stage~3 (right column) data estimated using the CLEAN algorithm with a
  gain of 0.05 and about 1400 iterations. The middle row corresponds to a
  magnified view of the 10 lowest modes from 0\--5~cycles/day for each stage
  corresponding to the small box in the lower left hand corner of the panels
  in the top row. The bottom row is a magnified view of the modes 650\--1000
  from 20\--30~cycles/day for each stage corresponding to the box in the lower
  near the middle of each panel in the top row.\label{fig:KL_SPECTRA}}
\end{figure}
The CLEAN algorithm \cite[]{Roberts1987,Hogbom1974} is used to estimate the
power spectral density of the KL-modes for each stage. The CLEAN algorithm
performs a nonlinear devolution of the unevenly temporally sampled
coefficients $\ALPHA$ in the frequency domain to attempt to minimize the
spectral artifacts introduced by the sampling
function. Figure~\ref{fig:KL_SPECTRA} shows the power spectral density in the
coefficients of the KL transforms $\left(\ALPHA^2\right)$ for stage 1 (left
column), stage 2 (middle column), and stage 3 (right column) data estimated
using the CLEAN algorithm with a gain of 0.05 and about 1400 iterations. . The
middle row is a magnified view of the 10 lowest modes from 0\--5~cycles/day
for each stage corresponding to the small box in the lower left hand corner of
the panels in the top row. The bottom row corresponds to a magnified view of
the modes 650\--1000 from 20\--30~cycles/day for each stage corresponding to
the box in the lower near the middle of each panel in the top row.\par
The general features of the top panel are a low frequency~0\--5~cycles/day
signature from modes 0\--100 with a narrow dispersion relationship $f\simeq\#$
above that. The low frequency signature is produced by large scale features
moving across the solar disk where as the narrow dispersion relation is caused
by small scale convective features moving across the disk. This narrow
dispersion relationship exhibits harmonics above mode 200.  The spectral
signatures of the orbital artifacts in the large scale flows is exhibited in
the middle panel of the first column. Significant isolated spectral peaks are
present at 1 and 2 cycles/day in mode~1 and at 1, 2, and 3 cycles/day in
mode~2. There are also harmonics at 1, 2, and 3 cycles/day near mode 400 in
the top left panel.  These spectral signatures are largely mitigated after the
removal of the large scale flow bias as shown by the panels in the middle and
right columns. However, what remains in the middle column is multiple
harmonics of the narrow dispersion relationship $f\simeq\#$. These harmonics
are cause by the modulation in the amplitude of convective structures as they
rotate across the disk. The stage 3 data shows that the spatially and
temporally dependent gain adjustment included in~(\ref{eqn:gain}) considerably
reduces the affect of this modulation by collapsing the harmonics to a single
peak at $f\simeq\#$. 
 
\makeatletter{}\section{Discussion and Conclusions}
 It is important to note that the stage 3 data cannot be claimed to be
 \textit{more accurate}, only that it is \textit{more consistent} from
 Dopplergram-to-Dopplergram. There are still artifacts which remain in the
 data shown in the right column of Figures~\ref{fig:KL_ALPHA}
 and~\ref{fig:KL_SPECTRA}, however, this is clearly a dramatic
 improvement in the quality of the Doppler data.  While it is well known that
 the Milne-Eddington inversion of the HMI Pipeline data contains orbital
 artifacts up to harmonics of several cycles/day \cite[]{Hoeksema2014}, the
 important take-away from the left hand column of Figures~\ref{fig:KL_ALPHA}
 and~\ref{fig:KL_SPECTRA} is that \textit{all temporal and spatial scales are
   contaminated by the orbital artifacts!} We know of no simple
 post-processing that will filter each spatial scale appropriately to remove
 the harmonics. It would be very surprising if the same conclusion did not
 also apply to the other critical observables produced by the Milne-Eddington
 inversion of the HMI Pipeline data such as the magnetic field
 data. Furthermore, if similar contamination is present in the 45 second
 Dopplergrams observed by Camera\#2, we speculate that the modulation of the
 observations will affect the amplitude of the 5 minute oscillations critical
 to Helioseismic observations.\par

 The important new result of this paper is that the \CODERED{}
 procedure does successfully remove the orbital artifacts in the HMI
 Doppler data. Figure~\ref{fig:raw_doppler} shows that the improvement
 in the data is dramatic; the daily oscillations are almost completely
 eliminated. Furthermore, the procedure is robust in that as shown by
 Figure~\ref{fig:KL_SPECTRA}, it cleans the data on all spatial scales
 without introducing new artifacts. The \CODERED{} procedure is
 straightforward to implement and will work on any data set of HMI
 Dopplergrams, consequently, we recommend that our procedure, or some
 modification, be incorporated into any HMI data analysis
 investigation. A key feature of the procedure is the use of the limb
 shift eigenfunctions to correct the gain in each pixel. The fact this
 gain correction is so successful has major implications for the
 possible mechanisms giving rise to the HMI errors and, consequently,
 for removing these errors from the vector magnetograms, as
 well. If we can correct the vector magnetograms to same level of fidelity
as that shown by Figures~\ref{fig:raw_doppler} and
~\ref{fig:KL_SPECTRA}, the resulting data would be invaluable for studying
solar coronal structure and dynamics.

\acknowledgments Peter Schuck acknowledges the support of the NASA LWS and GI
programs and NASA/GSFC bridge funding in completing this work. He also
gratefully acknowledges useful conversations with Peter Williams, Mark Linton,
Jesper Schou, David Hathaway, Yang Liu, Todd Hoeksema, and Phil
Scherrer. K.D. Leka, Graham Barnes, and Peter Schuck were supported by NASA GI
contracts NNH12CG10C: ``Photospheric Properties of Flaring vs. Flare-Quiet
Active Regions: Can we use HMI Vector Magnetogram Sequences to Quantify `When
and Why does the Sun go Boom?{'}'' and NNH12CC03C: ``Using SDO/HMI Data to
Investigate the Energization of the Coronal Magnetic Field.''  We also
gratefully acknowledge the help of Steve Martin who configured ``houdini'' our
80 core, 2TB shared memory system which was indispensable in the
calculations. We also acknowledge the use of GNU Parallel \cite[]{Tange2011}.

\appendix
\makeatletter{}\section{Observer Motion\label{app:observer}}
Using Heliocentric-Cartesian coordinates $\left(x,y,z\right)$, the $z$-axis is
defined along the axis parallel to the observer-Sun line, pointing towards the
observer. The $y$-axis is define to be perpendicular to that line and in the
plane with the $z$-axis and the solar North pole with $y$ increasing towards
solar North. The $x$-axis is defined to be perpendicular to both the $y$- and
$z$-axes with $x$ increasing towards solar West. The location of a feature on
the disk is given by \cite[see eq. (17) in][]{Thompson2006}
\begin{subequations}
\begin{align}
x=&-d\,\sin\theta_\rho\,\sin\psi,\label{eqn:helioprojective:x}\\
y=&d\,\sin\theta_\rho\,\cos\psi,\\
z=&D_\odot-d\,\cos\theta_\rho,
\end{align}
\end{subequations}
where $\theta_\rho$ is the helioprojective angle, $\psi$ is the position angle
defined counter-clockwise from solar North, $d$ is the distance between the
feature and the observer, and $D_\odot$ is the distance between the observer
and Sun center. The LOS vector pointing \textit{from} the feature \textit{to}
the observer is then
\begin{equation}
\boldsymbol{\widehat{\eta}_\mathrm{LOS}}\left(\theta_\rho,\psi\right)=\sin\theta_\rho\,\sin\psi\,\xhat-\sin\theta_\rho\,\cos\psi\,\yhat+\cos\theta_\rho\,\zhat.\label{eqn:eta:los}
\end{equation}
This implies that the raw observed Doppler velocity at each pixel (stage~0)
can be expressed as
\begin{subequations}
\begin{equation}
\vlos{0}=\boldsymbol{\widehat{\eta}_\mathrm{LOS}}\cdot\VSDO-\boldsymbol{\widehat{\eta}_\mathrm{LOS}}\cdot\boldsymbol{U}_\mathrm{surface},
\end{equation}
and the data with the satellite velocity subtracted or the ``V-sat
subtracted'' (stage~1) data may be expressed as
\begin{equation}
\vlos{1}=\vlos{0}-\boldsymbol{\widehat{\eta}_\mathrm{LOS}}\cdot\VSDO\equiv-\boldsymbol{\widehat{\eta}_\mathrm{LOS}}\cdot\boldsymbol{U}_\mathrm{surface}.\label{eqn:satellite}
\end{equation}
\end{subequations}
\section{Stonyhurst Unit Vectors\label{sec:Stony}}
Since we are describing the sun in Stonyhurst coordinates, the unit vectors
must be determined to resolve the projection of various vector quantities onto
the LOS. The transformation between Stonyhurst and heliocentric
Cartesian coordinates is
\begin{subequations}\label{eqn:Stonyhurst}
\begin{align}
x=&r\,\cos\Theta\,\sin\left(\Phi-\Phi_0\right),\\
y=&r\,\sin\Theta\,\cos{B_0}-\cos\Theta\,\cos\left(\Phi-\Phi_0\right)\,\sin{B_0},\\
z=&r\,\sin\Theta\,\cos{B_0}+\cos\Theta\,\cos\left(\Phi-\Phi_0\right)\,\cos{B_0},
\end{align}
\end{subequations}
where $\Theta$ and $\Phi$ are the latitude and longitude, and $B_0$ is the
so-called solar-B angle (the latitude of the center of the solar disk as seen
by the observer, and $\Phi_0$ is the Carrington longitude of the center of the
solar disk. The gradient in Stonyhurst coordinates is given by
\begin{equation}
\grad=\ephi\,\frac{1}{r\,\cos\Theta}\,\partial_\Theta+
\etheta\,\frac{1}{r}\,\partial_r+
\er\,\partial_r.
\end{equation}
The Jacobian of transformation between heliocentric Cartesian unit vectors and
Stonyhurst unit vectors is then
\begin{subequations}
\begin{align}
\Jacobian=&\grad\left(x,y,z\right),\\
&=\left(
                  \begin{array}{ccc}
                   \cos \Phi & \sin B_0 \sin \Phi & -\cos B_0 \sin \Phi \\
                   -\sin \Phi \sin \Theta & \cos B_0 \cos \Theta+\cos \Phi \sin B_0 \sin \Theta & \cos \Theta
                     \sin B_0-\cos B_0 \cos \Phi \sin \Theta \\
                   \cos \Theta \sin \Phi & \cos B_0 \sin \Theta-\cos \Phi \cos \Theta \sin B_0 & \cos B_0
                     \cos \Phi \cos \Theta+\sin B_0 \sin \Theta \\
                  \end{array}
                  \right)
\end{align}
\end{subequations}
where $\Jacobian$ transforms heliocentric Cartesian to Stonyhurst unit vectors
and $\Jacobian^\mathrm{T}$ vice-versa.
\begin{subequations}
\begin{align}
\left(\ephi,\etheta,\er\right)^\mathrm{T}=&\Jacobian\cdot\left(\xhat,\yhat,\zhat\right)^\mathrm{T},\\
\left(\xhat,\yhat,\zhat\right)^\mathrm{T}=&\Jacobian^\mathrm{T}\cdot\left(\ephi,\etheta,\er\right)^\mathrm{T}.
\end{align}
\end{subequations}
These unit vectors can be combined with~(\ref{eqn:eta:los}) to determine the
projection of the Stonyhurst unit vectors onto the LOS.
\begin{subequations}
\begin{align}
\boldsymbol{\widehat{\eta}_\mathrm{LOS}}\cdot\Jacobian^\mathrm{T}\cdot\ephi=& -\cos B_0\,\sin\Phi\, \cos\theta_\rho+\left(\cos \Phi\,\sin \psi-\sin B_0\,\sin \Phi\,\cos \psi\right)\,\sin \theta_\rho,\\ 
\boldsymbol{\widehat{\eta}_\mathrm{LOS}}\cdot\Jacobian^\mathrm{T}\cdot\etheta=&\left(\sin
B_0 \,\cos \Theta-\cos B_0 \,\cos \Phi \,\sin \Theta\right)\,\cos \theta_\rho\\
&-\left[\sin \Phi\,\sin \Theta\,\sin\psi+(\sin B_0\,\cos \Phi \,\sin \Theta+\cos B_0 \,\cos \Theta)\,\cos \psi\right]\,\sin \theta_\rho\nonumber\\ 
\boldsymbol{\widehat{\eta}_\mathrm{LOS}}\cdot\Jacobian^\mathrm{T}\cdot\er=&\left(\cos
B_0\,\cos \Phi\,\cos \Theta+\sin B_0\,\sin \Theta\right)\,\cos\theta_\rho\\ &+\left[\sin \Phi\,\cos \Theta\,\sin \psi -\left(\cos B_0 \,\sin \Theta-\sin B_0 \,\cos
    \,\Phi \cos \Theta\right)\,\cos \psi\right]\,\sin \theta_\rho.\nonumber
\end{align}
\end{subequations}
\section{Heliocentric Coordinates for Meridional Flows and the Convective Blue-Shift\label{sec:Helio}}
The convective blue-shift is conventionally described in heliocentric
spherical coordinates where $\varrho$ is the angle between the point on the
solar surface and the line connecting Sun center to the observer ($\zhat$\--axis).
\begin{subequations}\label{eqn:heliocentric:spherical}
\begin{align}
x=&-R_\odot\,\sin\varrho\,\sin\psi,\label{eqn:sphere:x}\\
y=&R_\odot\,\sin\varrho\,\cos\psi,\\
z=&R_\odot\,\cos\varrho,
\end{align}
\end{subequations}
Equating~(\ref{eqn:helioprojective:x}) and~(\ref{eqn:sphere:x}) we can obtain
the law sines for the relationship between the angles
\begin{equation}
\frac{\sin\varrho}{d}=\frac{\sin\theta_\rho}{R_\odot}=\frac{\sin\left[\pi-\left(\varrho+\theta_\rho\right)\right]}{D_\odot},
\end{equation}
where the last relationship is determined from the law of sines.
Rearranging we have \cite[see pp. 174--175 in][]{Smart1977} 
\begin{subequations}
\begin{equation}
\sin\left(\varrho+\theta_\rho\right)=\frac{D_\odot}{R_\odot}\,\sin\theta_\rho
\end{equation}
or
\begin{equation}
\varrho=\sin^{-1}\left(\frac{D_\odot}{R_\odot}\,\sin\theta_\rho\right)-\theta_\rho.
\end{equation}
\end{subequations}
Equation~(\ref{eqn:Stonyhurst}) and~(\ref{eqn:heliocentric:spherical}) we
obtain the relations between Stonyhurst and heliocentric spherical coordinates
\begin{subequations}
\begin{align}
\sin\Theta=&\sin B_0\,\cos \varrho+ \cos B_0\,\sin \varrho\, \cos \psi ,\\
\cos\Theta\,\sin\Phi=&-\sin\varrho\,\sin\psi,\\
\cos\Theta\,\cos\Phi=& \cos B_0 \,\cos \varrho-\sin B_0\,\sin \varrho\, \cos \psi.
\end{align}
\end{subequations}
\section{Evaluation of Hathaway's Integral\label{sec:GMF}}
\begin{equation}
G_{\mathrm{MF},\ell}\left(B_0,\varrho\right)=\frac{1}{2\,\pi}\,\int_0^{2\,\pi}d\psi\,\boldsymbol{\widehat{\eta}_\mathrm{LOS}}\cdot\Jacobian^\mathrm{T}\cdot\etheta\,\Pbar_\ell^1\left(\sin\Theta\right).\label{eqn:GMF}
\end{equation}
This is a complicated integral to evaluate given that $\Theta$ and $\Phi$ must
be re-expressed as functions of $\varrho$ and $\psi$. Noting that the
associated Legendre function may be expressed in terms of the Legendre
function of order zero $\left(m=0\right)$
\begin{equation}
\Leg_\ell^m\left(x\right)=\left(-1\right)^m\,\left(1-x^2\right)^{m/2}\,\frac{d^m}{dx^m}\Leg_\ell\left(x\right),
\end{equation}
and using a Lemma for the
expansion of the derivative of Legendre functions $\Leg_\ell^m\left(x\right)$ \cite[See][]{Garfinkel1964}
\begin{equation}
\frac{d\Leg_\ell\left(x\right)}{dx}=\sum_{n=0}^{\left(\ell-1\right)/2}\left(2\,\ell-4\,n-1\right)\,\Leg_{\ell-1-2\,n}\left(x\right),
\end{equation}
the associated Legendre function of order one may be expressed as a
terminating series of Legendre functions of order zero
\begin{equation}
\Leg_\ell^1\left(x\right)=-\sqrt{1-x^2}\,\sum_{n=0}^{\left(\ell-1\right)/2}\left(2\,\ell-4\,n-1\right)\,\Leg_{\ell-1-2\,n}\left(x\right).
\end{equation}
 Substituting
$x=\sin\Theta$ 
\begin{equation}
\Leg_\ell^1\left(\sin\Theta\right)=-\cos\Theta\,\sum_{n=0}^{\left(\ell-1\right)/2}\left(2\,\ell-4\,n-1\right)\,\Leg_{\ell-1-2\,n}\left(\sin\Theta\right),
\end{equation}
and using the Spherical Harmonic Addition Theorem
\begin{align}
\Leg_\ell\left(\sin\Theta\right)=&\frac{4\,\pi}{2\,\ell+1}\,\sum_{m=-\ell}^\ell{Y_\ell^m\left(B_0,\psi/2\right)\,Y_\ell^{m*}\left(\varrho,-\psi/2\right)},\nonumber\\
=&\frac{2}{2\,\ell+1}\,\sum_{m=-\ell}^\ell
\Pbar_\ell^m\left(\sin{B_0}\right)\,
\Pbar_\ell^m\left(\cos\varrho\right)\,e^{i\,m\,\psi},
\end{align}
this becomes
\begin{equation}
\Leg_\ell^1\left(\sin\Theta\right)=-2\,\cos\Theta\,\sum_{n=0}^{\left(\ell-1\right)/2}\,\sum_{m=-\left({\ell-1-2\,n}\right)}^{\ell-1-2\,n}
\Pbar_{\ell-1-2\,n}^m\left(\sin{B_0}\right)\,
\Pbar_{\ell-1-2\,n}^m\left(\cos\varrho\right)\,e^{i\,m\,\psi}
\end{equation}
where $\psi$ is now external to the Legendre functions.
Using~(\ref{eqn:Pbar})
\begin{equation}
\Pbar_\ell^1\left(x\right)=-\sqrt{\frac{\left(2\,\ell+1\right)\,\left(\ell-1\right)!}{2\,\left(\ell+1\right)!}}\,\Leg_\ell^1\left(x\right),
\end{equation}
we obtain a form consistent with~(\ref{eqn:GMF})
\begin{equation}
\Pbar_\ell^1\left(\sin\Theta\right)=2\,\cos\Theta\,\sqrt{\frac{\left(2\,\ell+1\right)\,\left(\ell-1\right)!}{2\,\left(\ell+1\right)!}}\,\sum_{n=0}^{\left(\ell-1\right)/2}\,\sum_{m=-\left({\ell-1-2\,n}\right)}^{\ell-1-2\,n}
\!\!\!\!\!\!\Pbar_{\ell-1-2\,n}^m\left(\sin{B_0}\right)\,
\Pbar_{\ell-1-2\,n}^m\left(\cos\varrho\right)\,e^{i\,m\,\psi}
\end{equation}
Thus the integral may be evaluated as
\begin{align}
G_{\mathrm{MF},\ell}\left(B_0,\varrho\right)=&2\,\sqrt{\frac{\left(2\,\ell+1\right)\,\left(\ell-1\right)!}{2\,\left(\ell+1\right)!}}\sum_{n=0}^{\left(\ell-1\right)/2}\,\sum_{m=-\left({\ell-1-2\,n}\right)}^{\ell-1-2\,n}
\!\!\!\!\!\!\Pbar_{\ell-1-2\,n}^m\left(\sin{B_0}\right)\,
\Pbar_{\ell-1-2\,n}^m\left(\cos\varrho\right)\nonumber\\
&\times\,\left[I_1^m\left(B_0,\rho\right)\,\cos \theta_\rho-I_2^m\left(B_0,\rho\right)\,\sin \theta_\rho\right].
\end{align}
where
\begin{subequations}
\begin{align}
I_1^m\left(B_0,\rho\right)=&\frac{1}{2\,\pi}\int_0^{2\,\pi}\,d\psi\,e^{i\,m\,\psi}\,\cos\Theta\,\left(\sin B_0 \,\cos \Theta-\cos B_0 \,\sin \Theta\,\cos \Phi\right),\\
I_2^m\left(B_0,\rho\right)=&\frac{1}{2\,\pi}\int_0^{2\,\pi}\,d\psi\,e^{i\,m\,\psi}\,\cos\Theta\,\left[\sin \Theta\,\sin \Phi\,\sin
    \psi\right.\nonumber\\ &\qquad\qquad\qquad+\left.\left(\sin B_0 \,\sin \Theta\,\cos \Phi +\cos B_0 \,\cos
  \Theta\right)\,\cos
  \psi\right]\nonumber.
\end{align}
Evaluating the first integral
\end{subequations}
\begin{subequations}
\begin{align}
I_1^m\left(B_0,\varrho\right)=&\frac{1}{2\,\pi}\,\int_0^{2\,\pi}d\psi\,e^{i\,m\,\psi}\,\cos\Theta\,\left(\sin{B_0}\,\cos\Theta-\cos{B_0}\,\sin\Theta\,\cos\Phi\right),\nonumber\\
=&\frac{\sin\varrho}{2\,\pi}\,\int_0^{2\,\pi}d\psi\,e^{i\,m\,\psi}\,\left(\sin{B_0}\,\sin\varrho-\cos{B_0}\,\cos\varrho\,\cos\psi\right),\nonumber\\
=&(-1)^{m+1}\,\sin\varrho\,\left[\left(2 m^2-1\right)
  \cos\left(B_0+\varrho\right)+\cos\left(B_0-\varrho\right)\right]\,\frac{\sin\left(m\,\pi\right)\,}{2
  \pi  m \left(m^2-1\right)},\nonumber\\
=&\delta_{m,0}\,\sin{B_0}\,\sin^2\varrho-\frac{1}{2}\,\delta_{\left|m\right|,1}\,\cos{B_0}\,\sin\varrho\,\cos\varrho.
\end{align}
Similarly
\begin{equation}
I_2^m\left(B_0,\varrho\right)=\delta_{m,0}\,\sin{B_0}\,\sin\varrho\,\cos\varrho-\frac{1}{2}\,\delta_{\left|m\right|,1}\,\cos{B_0}\,\cos^2\varrho.
\end{equation}
\end{subequations}
Noting that $I_2^m\left(B_0,\varrho\right)=-\cot\varrho\,I_1^m\left(B_0,\varrho\right)$
\begin{align}
G_{\mathrm{MF},\ell}\left(B_0,\varrho\right)=&\left(\cos
\theta_\rho+\cot\varrho\,\sin \theta_\rho\right)\\
&\times2\,\sqrt{\frac{\left(2\,\ell+1\right)\,\left(\ell-1\right)!}{2\,\left(\ell+1\right)!}}\,\sum_{m=-1}^{1}I_1^m\left(B_0,\varrho\right)\sum_{n=0}^{\left(\ell-1\right)/2}\Pbar_{\ell-1-2\,n}^m\left(\sin{B_0}\right)\,
\Pbar_{\ell-1-2\,n}^m\left(\cos\varrho\right).\nonumber
\end{align}
Noting the symmetry property of the Legendre functions
\begin{subequations}
\begin{align}
\Leg_\ell^{-m}\left(x\right)=&\left(-1\right)^m\,\frac{\left(\ell-m\right)!}{\left(\ell+m\right)!}\,\Leg_\ell^m\left(x\right),\\
\Pbar_\ell^{-m}\left(x\right)=&\left(-1\right)^m\,\Pbar_\ell^m\left(x\right).
\end{align}
\end{subequations}
\begin{align}
G_{\mathrm{MF},\ell}\left(B_0,\varrho\right)=&\sqrt{\frac{2}{2\,\ell+1}}\,\sin\left(\varrho+\theta_\rho\right)\,\sqrt{1+2\,\ell}\,\sqrt{\frac{2\,\ell+1}{\ell\,\left(\ell+1\right)}}\times\\
&\sum_{n=0}^{\left(\ell-1\right)/2}\left[\Pbar_{\ell-1-2\,n}^0\left(\sin{B_0}\right)\,
\Pbar_{\ell-1-2\,n}^0\left(\cos\varrho\right)\,\sin{B_0}\,\sin\varrho\right.\nonumber\\
&\qquad\qquad\left.-\Pbar_{\ell-1-2\,n}^1\left(\sin{B_0}\right)\,
\Pbar_{\ell-1-2\,n}^1\left(\cos\varrho\right)\,\cos{B_0}\,\cos\varrho\right].\nonumber
\end{align}
Employing an inductive proof this generally becomes 
\begin{align}
G_{\mathrm{MF},0}\left(B_0,\varrho\right)=&0,\\
G_{\mathrm{MF},1}\left(B_0,\varrho\right)=&\sqrt{\frac{2}{3}}\,\sin\left(\varrho+\theta_\rho\right)\,\frac{3}{2\,\sqrt{2}}\,\sin{B_0}\,\sin{\varrho},\nonumber\\
=&\sqrt{\frac{2}{3}}\,\sin\left(\varrho+\theta_\rho\right)\,\Pbar_1^0\left(\sin{B}_0\right)\,\Pbar_1^1\left(\cos\varrho\right),\\
G_{\mathrm{MF},2}\left(B_0,\varrho\right)=&\sqrt{\frac{2}{5}}\,\sin\left(\varrho+\theta_\rho\right)\,\frac{5}{8}\,\sqrt{\frac{3}{2}}\,\left[1-3\,\cos\left(2\,B_0\right)\right]\,\sin{\varrho}\,\cos\varrho,\nonumber\\
=&\sqrt{\frac{2}{5}}\,\sin\left(\varrho+\theta_\rho\right)\,\Pbar_2^0\left(\sin{B}_0\right)\,\Pbar_2^1\left(\cos\varrho\right),
\end{align}
and generally
\begin{equation}
G_{\mathrm{MF},\ell}\left(B_0,\varrho\right)=\sqrt{\frac{2}{2\,\ell+1}}\,\sin\left(\varrho+\theta_\rho\right)\,\Pbar_\ell\left(\sin{B_0}\right)\,\Pbar_\ell^1\left(\cos\varrho\right).
\end{equation}
where $\theta_\rho\approx0$ corresponds to the results of \cite{Hathaway1992}.
\subsection{Karhunen-Lo\'eve (KL) Analysis\label{sec:KL}}
The Karhunen-Lo\'eve (KL) Analysis \cite[]{Loeve1955} is known by various
names: Principle Component Analysis, Proper Orthogonal Decomposition,
Empirical Orthogonal Functions, and/or the Hoteling transform. The approach
presented here follows the Method of Snapshots
\cite[]{Lumley1967,Sirovich1987,Holmes1996}. The goal of KL is to determine an
optimal representation for the data \--- optimal in the sense that the vector
$\KLeig$ maximizes the projection onto the median subtracted
$\REAL{\ND}{\Nt}$ image array $\IS$. This optimality can be expressed by
the functional
\begin{equation}
\mathcal{J}\left(\KLeig\right)=\ND^{-1}\,\KLeig^\dag\,\IS^\dag\,\IS\,\KLeig-\lambda\,\left(\KLeig^\dag\cdot\KLeig-1\right),
\end{equation}
where $\dag$ represents the conjugate transpose for complex data or just the
transpose for real data as is being considered here and the constraint ensures
normalization.
The $\COMPLEX{\NT}{\NT}$ spatially averaged covariance matrix of the image sequence
is defined as
\begin{subequations}
\label{eqn:autocovariance}
\begin{align}
\ACM&\equiv\frac{\IS^\dag\,\IS}{\ND},\\
\acm_{\Index{m}}&=\frac{1}{\ND}\,\sum_{i=1}^{\ND}{\left[\IS^\dag\right]}_{\Index{i}}\,\left[\IS\right]_{im},
\end{align}
\end{subequations}
which permits the expression of the functional
\begin{equation}
\mathcal{J}\left(\KLeig+\delta\deig\right)=
\left(\KLeig^\dag+\delta\deig^\dag\right)\,\ACM\,\left(\KLeig+\delta\deig\right)-\lambda\,\left[\left(\KLeig^\dag+\delta\deig^\dag\right)\,\left(\KLeig+\delta\deig\right)-1\right]
\end{equation}
where $\delta\deig$ is a small vector perturbation on $\KLeig$. 
Using the calculus of variations, the first variation of the functional is
\begin{subequations}
\begin{align}
\lim_{\delta\rightarrow0}\frac{d}{d\delta}\mathcal{J}\left(\KLeig+\delta\deig\right)
&=\deig^\dag\,\ACM\,\KLeig+\KLeig^\dag\,\ACM\,\deig-\lambda\,\left(\deig^\dag\,\KLeig+\KLeig^\dag\,\deig\right),\\
&=\deig^\dag\,\left(\ACM\,\KLeig-\lambda\,\KLeig\right)+\left(\KLeig^\dag\,\ACM^\dag-\lambda\,\KLeig^\dag\right)\,\deig,\\
&=\deig^\dag\,\left(\ACM\,\KLeig-\lambda\,\KLeig\right)+\left[\deig^\dag\,\left(\ACM\,\KLeig-\lambda\,\KLeig\right)\right]^\dag.
\end{align}
\end{subequations}
Therefore, finding the vectors which maximize they projection onto the data is
equivalent to finding the eigenvalues $\lambda_k$ and eigenvectors $\KLeig_k$
of the covariance matrix.
\begin{subequations}
\label{eqn:eigenequation}
\begin{equation}
\ACM\,\KLeig_k=\KLeig_k\,\lambda_k,
\end{equation}
or
\begin{equation}
\ACM\,\KLEIG=\KLEIG\,\LAMBDA,
\end{equation}
\end{subequations}
where the columns of
$\KLEIG=\left(\KLeig_1,\KLeig_2,\ldots,\KLeig_{\NT}\right)$ are the
eigenvectors of $\ACM$ and $\LAMBDA$ is a diagonal matrix of the eigenvalues
in decreasing order. The covariance in~(\ref{eqn:autocovariance}) is a
hermitian $\ACM=\ACM^\dag$ positive semi-definite matrix with non-negative
eigenvalues that can be ordered by decreasing value
$\lambda_1\ge\lambda_2\ge...\ge\lambda_{\NT-1}\ge\lambda_{\NT}\ge0$ with
orthonormal eigenvectors\footnote{Since the temporal median of each pixel is
  subtracted from the image sequence, the last eigenvalue $\lambda_{\NT}$ will
  often be close to zero. Precision errors can manifest themselves as negative
  eigenvalues despite the positive semi-definite properties of the matrix
  $\ACM$. Standard practice is to set this negative eigenvalue to zero
  effectively ignoring it in any modal reconstruction.}
\begin{subequations}
\begin{align}
\delta_{ij}&=\KLeig_i^\dag\cdot\KLeig_j,\\
\delta_{ij}&=\sum_{\Index=1}^{\NT}\DeltaD_{\Index{i}}^*\,\DeltaD_{\Index{j}},\\
\Identity&=\KLEIG^\dag\,\KLEIG,\label{eqn:orthogonality}
\end{align}
\end{subequations}
where $\Identity$ is the identity matrix. Assuming that the image-sequence can
be described by a $\COMPLEX{\ND}{\NT}$ matrix of spatial eigenmodes $\PHI$ and
$\COMPLEX{\NT}{\NT}$ matrix of temporal coefficients $\ALPHA$, this
\textit{ansatz} takes the form
\begin{subequations}
\begin{equation}
\IS=\PHI\,\ALPHA^\dag,\label{eqn:galerkin}
\end{equation}
\begin{equation}
\ALPHA\equiv\KLEIG\,\LAMBDA^{1/2}.
\end{equation}
\end{subequations}
Post-multiplying~(\ref{eqn:galerkin}) by $\KLEIG$ and using the orthogonality relationship~(\ref{eqn:orthogonality})
\begin{equation}
\IS\,\KLEIG=\PHI\,\LAMBDA^{1/2}
\end{equation}
results in the equation for the spatial eigenfunctions $\PHI$ where
\begin{equation}
\PHI\equiv\IS\,\KLEIG\,\LAMBDA^{-1/2}
\end{equation}
Multiplying this by its adjoint 
\begin{equation}
\PHI^\dag\,\PHI=\LAMBDA^{-1/2}\,\KLEIG^\dag\,\IS^\dag\,\IS\,\KLEIG\,\LAMBDA^{-1/2},
\end{equation}
and using~(\ref{eqn:autocovariance})
\begin{equation}
\PHI^\dag\,\PHI=\ND\,\LAMBDA^{-1/2}\,\KLEIG^\dag\,\ACM\,\KLEIG\,\LAMBDA^{-1/2},
\end{equation}
followed by~(\ref{eqn:eigenequation}) produces
\begin{equation}
\PHI^\dag\,\PHI=\ND\,\LAMBDA^{-1/2}\,\KLEIG^\dag\,\KLEIG\,\LAMBDA\,\LAMBDA^{-1/2},
\end{equation}
which by virtue of~(\ref{eqn:orthogonality}) becomes the orthogonality
relationship for the spatial eigenfunctions
\begin{equation}
\PHI^\dag\,\PHI=\ND\,\LAMBDA^{-1/2}\,\LAMBDA\,\LAMBDA^{-1/2}=\ND\,\Identity.
\end{equation}
The temporal coefficients are orthogonal
\begin{equation}
\ALPHA^\dag\,\ALPHA=\LAMBDA,
\end{equation}
which implies that the temporal dynamics of the spatial eigenfunctions is on
average uncorrelated. The temporal coefficients may also be used to reconstruct
the covariance matrix
\begin{equation}
\ACM=\ALPHA\,\ALPHA^\dag.
\end{equation}
An analogous methodology could be applied to the time-averaged
two-point time-averaged spatial correlation function $\IS\IS^\dag/\NT$ instead
of the spatially averaged correlation function
in~(\ref{eqn:autocovariance}). However, this direct method leads to an
$\COMPLEX{\ND}{\ND}$ correlation matrix with $\ND\simeq10^7$ spatial
eigenvalues and eigenvectors which is prohibitively large for present
computers. The method of snapshots, described above, is a practical method to
obtain similar results when $\NT\ll\ND$ \cite[]{Sirovich1987}.
 
\makeatletter{}

\clearpage

\end{document}